\documentstyle[psfig]{l-aa}
\newcommand{\ha}{H$\alpha~$}

\newcommand{\etal}{{\it et al. }}
\newcommand{\kms}{km~s$^{-1}~$}

\newcommand{\msol}{M$_{\odot}$~} %\newcommand{\msol}{M$_{\odot}~$}
\newcommand{\lsol}{L$_{\odot}$~} %\newcommand{\lsol}{L$_{\odot}~$}

\newcommand{\pyr}{yr$^{-1}$}
\newcommand{\mum}{$\mu$m~}

\newcommand{\Ctwo}{\mbox{C$_{2}$}}
\newcommand{\Cthree}{\mbox{C$_{3}$}}
\newcommand{\CHp}{\mbox{CH$^{+}$}}
\newcommand{\Htwo}{\mbox{H$_{2}$}}
\newcommand{\philband}{$\rm A^{1}\Pi_{u}-X^{1}\Sigma^{+}_{g}~$}
\newcommand{\redband}{$\rm A^{2}\Pi-X^{2}\Sigma^{+}~$}
\newcommand{\swanband}{$\rm d^{3}\Pi_{g}-a^{3}\Pi_{u}~$ }
\newcommand{\Cthreeband}{$\rm A^{1}\Pi_{u}-X^{1}\Sigma_{g}^{+}~$}

\begin{document}
\thesaurus{08(          % A&A section 8: Stars
           02.12.2;     % line: identification
           02.13.4;     % molecular data
           02.13.5;     % molecular processes
           08.03.4;     % circumstellar matter
           08.09.2;     % stars: individual: HD~56126
           08.16.4)}    % stars: AGB and post-AGB
\title{Detection of \Ctwo, CN, and NaI~D
      absorption in the AGB remnant of HD~56126
       \thanks{Based on observations with the Utrecht Echelle
                 Spectrograph on the William Herschel Telescope
                 (La Palma, Spain)}
      \thanks{Tables of App.~A are only available in electronic form
              at the CDS via anonymous ftp 130.79.128.5}}
\author{Eric J. Bakker\inst{1,2}\and
        L.B.F.M. Waters\inst{3,4}\and
        Henny J.G.L.M. Lamers\inst{2,1}\and
        Norman R. Trams\inst{5}\and
        Frank L.A. Van der Wolf\inst{1}}
\offprints{Eric J. Bakker, present address:
        Astronomy Department,
        University of Texas, RLM 16.218,
        Austin, TX 78712,
        U.S.A., ebakker@astro.as.utexas.edu}
\institute{Astronomical Institute, University of Utrecht,
           P.O. Box 80000,
           NL-3508 TA Utrecht,
           The Netherlands
           \and
           SRON Laboratory for Space Research,
           Sorbonnelaan 2,
           NL-3584 CA Utrecht,
           The Netherlands
           \and
           Astronomical Institute,
           University of Amsterdam,
           Kruislaan 403,
           NL-1098 SJ Amsterdam,
           The Netherlands
           \and
           SRON Laboratory for Space Research,
           P.O. Box 800,
           NL-9700 AV Groningen,
           The Netherlands
           \and
           ESTEC,
           P.O. Box 299,
           NL-2200 AG Noordwijk (ZH),
           The Netherlands}
\date{Received January 20 1995, accepted September 1995}
\maketitle

\begin{abstract}
We present the detection of molecular absorption lines in the
optical spectrum of the post-AGB star HD~56126. The \Ctwo~
Phillips $\rm A^{1}\Pi_{u}-X^{1}\Sigma^{+}_{g}$
(1,0), (2,0), and (3,0); Swan $\rm d^{3}\Pi_{g}-a^{3}\Pi_{u}$
(0,0) and (1,0); and CN Red system $\rm A^{2}\Pi-X^{2}\Sigma^{+}$
(1,0), (2,0), (3,0), and (4,0) bands have been identified.
{}From the identification of the molecular bands we find an expansion
velocity of $8.5\pm0.6$~\kms independent of excitation condition or
molecular specie.
On the basis of the expansion velocity,
rotational temperatures, and molecular column densities
we argue that the line-forming region is the AGB remnant.
This is in agreement with the expansion velocity derived from the
CO lines. We find column densities of
$\log N_{\rm C_{2}}=15.3\pm0.3$~cm$^{-2}$ and
$\log N_{\rm CN  }=15.5\pm0.3$~cm$^{-2}$, and
rotational temperatures of
$T_{\rm rot}=242\pm20$~K and $T_{\rm rot}=24\pm5$~K respectively for
C$_{2}$ and CN.

By studying molecular line absorption in optical spectra of post-AGB
stars we have found a new tracer of the AGB remnant.
{}From comparison with
the results of CO and IR observations it is possible to obtain information
on non-spherical behavior of the AGB remnant. Using
different molecules with different excitation conditions it should
be possible to study the AGB remnant as a function of the distance to the
star, and thus as a function of the evolutionary status of the star
on the AGB.
\end{abstract}

\section{Introduction}

After the IRAS mission a large number of studies have been conducted in
selecting post-Asymptotic Giant Branch (post-AGB)
stars based on the IRAS point source catalogue using
selection criteria on the infrared colors
(e.g., Trams~\etal \cite{art4tramsetalcat};
Oudmaijer~\etal \cite{art4oudmaijeretalcat}). By looking at the infrared
excess one can select sources on its dust geometry
(dust density and temperature distribution) rather than on evolutionary
status. To be certain about the post-AGB nature one needs additional
constraints. HD~56126 not only shows a characteristic infrared
excess for post-AGB stars, but also  many absorption lines
from s-process material.
S-process material is only dredged-up to the surface of
intermediate-mass stars during the AGB phase (Iben \cite{art4iben}).

During the end of the AGB phase the star suffers from an extreme mass
loss rate of
typically $\dot M_{\rm AGB} =10^{-5}$~\msol~\pyr, and the material
moves away from the central star with a typical velocity
of $v_{\rm exp}=10~{\rm to}~20$~km~s$^{-1}$.
The dust in the AGB remnant is in radiative
equilibrium
with the radiation field coming from the star: it absorbs UV
and optical radiation
and emits the energy in the far-IR. This causes a double-peaked energy
distribution,  one peak in the optical
from the stellar radiation and a second peak in the infrared
from the dust shell. The geometry of the mass loss
--- spherically symmetric or axisymmetric --- on the AGB
determines the geometry of the dust shell and thus
the amount of extinction the central star will suffer
as a post-AGB star.

HD~56126 has been classified as F5I
(Nassau~\etal \cite{art4nassauetal}) with a strong infrared source
peaking at 25~\mum (112.8 Jansky; IRAS PSC \cite{art4iraspsc}).
In the HD catalogue (published in 1919)
HD~56126 is classified as a G5 star, which
means that the star has possibly evolved from G5 to F5 in about 50 years.

Knowledge of the mass-loss rate on the AGB and the physical conditions of
the AGB remnant are of eminent importance in understanding the AGB and
post-AGB evolution of low- and intermediate-mass stars.
The amount of mass loss on the AGB seriously affects (i.e., dominates)
the mass in the
convective envelope of the AGB star. When the
envelope  mass decreases and reaches a critical value
of about 10$^{-2}$~M$_{\odot}$, the star
leaves the AGB and moves along a constant luminosity track to the White
Dwarf (WD) phase
(Sch\"{o}nberner \cite{art4schonberner83}).

Until now, studies of the AGB remnant of HD 56126 were
mainly  concentrated on
the millimeter and radio line emission of molecules,
e.g., $^{12}$CO, $^{13}$CO,
HCN (Omont~\etal \cite{art4omont}),
and the tentative detection of HCO$^{+}$
(Bujarrabal~\etal  \cite{art4bujarrabaletal92};
Nyman~\etal \cite{art4nymanetal92}),
as well as the infrared  excess
(Parthasarathy and Pottasch \cite{art4parthasarathypottasch}) and
dust features (Kwok~\etal \cite{art4kwoketal89}).
Although these are powerful tracers of the AGB remnant,
they all face the same problem that these tracers are in emission and
that spatial information (projected angle of the velocity vector)
and velocity information (expansion velocity) cannot be
determined independently if the source is not spatially resolved.

Here we present the detection of molecular absorption lines
in the spectrum of
the AGB remnant of HD~56126. This allows us to determine
outflow velocities, rotational temperatures, column densities, and
relative abundances in
the AGB remnant at a well-defined position
(Bakker~\etal \cite{art4bakkeredin}):
in the line of sight to the post-AGB star.
Comparison of
the results of this work with the results from studies on the emission
of the AGB remnant enables us to study the asymmetry of the AGB remnant
and mass-loss rates as a function of evolutionary status when the
star was still on the AGB.

Here we will first study the optical spectrum of HD~56126.
Sect.~2 gives information on how the spectra were obtained and reduced.
In Sect.~3 we present the  detection of
molecular absorption (\Ctwo~ and CN) in the AGB remnant.
In Sect.~4 we look in detail at  the physical conditions of the
line-forming region of \Ctwo~ and CN,
while in Sect.~5 we concentrate on the observed velocities and
predict the expected velocities for interstellar lines.
We discuss (Sect.~6) these observations
in terms of AGB mass loss and evolutionary time scales and
summarize the results in Sect.~7.
App.~\ref{art4ap-mollist} (only available at CDS) gives a line
identification list of the molecular lines of the
Phillips \Ctwo~ (1,0), (2,0), and (3,0) bands
and the CN Red system (1,0), (2,0), and (3,0).
App.~\ref{art4ap-atomlist}
gives the identification of selected lines of CI, NI, and OI.

\section{Observations and reduction of the echelle spectra}

The observations of HD~56126 (Table~\ref{art4tab-star})
were made on February $24^{{\rm th}}$ 1992
by HJGLM using two different grating settings of
echelle E31 with the Utrecht Echelle
Spectrograph (UES) on the William Herschel Telescope (WHT) on La Palma.
The spectra were obtained using the  EEV3 detector which has
$1180\times1180~22\left( \mu{\rm m} \right) ^{2}$ pixels.
The wavelength coverage of the spectrum is
from 4430 to 10302~\AA, but inherent to an echelle spectrograph
there are wavelength parts in the red ($\lambda > 6030 $~\AA)
which are not covered by the CCD
because the projected echellogram is larger than the size of the CCD detector.
The spectra were reduced by
Ton Schoenmaker in Roden using the IRAF package:
bias subtracted, wavelength calibrated, and continuum corrected.
The wavelength calibration was made using a ThAr arc spectrum and a
wavelength identification table of 1838 ThAr lines in
wavelength range between 3200 and 9548~\AA. For wavelengths longer than
9548~\AA~ the wavelength calibration was extrapolated using the fit
derived from  shorter wavelengths.
A log of the spectra is given  in Table~\ref{art4tab-obs}.
The velocity resolution  was determined from
the identified telluric lines
(Moore~\etal ~\cite{art4moore66}). The FWHM spread is
$\approx1.0$~km~s$^{-1}$.  At \ha (6563~\AA) the signal-to-noise-ratio is
$SNR=100$.
All velocities given in this article are heliocentrically corrected using
the correction given by the MIDAS routine COMPUTE/BARY
($\delta v_{\oplus}$ in Table~\ref{art4tab-obs}).

\begin{table}
\caption{Observational parameters of HD~56126 (SAO~96709, IRAS~07134+1005) }
\label{art4tab-star}
\centerline{\begin{tabular}{lll}
\hline
             &                                    &Remark
                    \\
\hline
             &                                    &
                     \\
Sp.T         & F5I                                &Nassau~\etal
\cite{art4nassauetal}                \\
Sp.T         & G5                                 &HD catalogue (1919)
                     \\
$T_{\rm eff}$& $6500$~K                           &Parthasarathy~\etal
\cite{art4parthasarathyetal92}\\
$\log g$     & $-1.4$                             &Parthasarathy~\etal
\cite{art4parthasarathyetal92}\\
$\alpha$     & $07^{\rm h}13^{\rm m}25.30^{\rm s}$&B1950
                \\
$\delta$     & $+10^{\rm o}05^{\rm '}09^{\rm ``}$ &B1950
                \\
$l^{\rm II}$ & $206.75^{\rm o}$                   &
                \\
$b^{\rm II}$ & $ +9.99^{\rm o}$                   &
                \\
$V$          & $8.23$                             &Hrivnak~\etal
\cite{art4hrivnaketal89}       \\
$B$          & $9.13$                             &Hrivnak~\etal
\cite{art4hrivnaketal89}       \\
$(B-V)_{o}$  &$0.30$                            &F5I                      \\
$E(B-V)$     &$0.60$                              &                         \\
$v_{\rm *CO}$&$85.6\pm1.0$~\kms                   &CO millimeter emission   \\
$\delta v_{\odot}$&$-14.11$~\kms              &helio $\rightarrow$ lsr  \\
             &                            &                         \\
\hline
\end{tabular}}
\end{table}

The conversion between different coordinate systems is made according
to Eq.~\ref{art4eq-velcorlsr} and \ref{art4eq-velcorearth},
where $v_{\oplus}$ is the observed velocity and $\delta v_{\oplus}$
the correction term to the heliocentric coordinate system ($v_{\odot}$).
Adding an extra term of $\delta v_{\odot}$,gives the  velocity in the
Local Standard of Rest system  ($v_{\rm lsr}$).

\begin{eqnarray}
\label{art4eq-velcorlsr}
v_{\odot}=v_{\oplus} + \delta v_{\oplus} \\
\label{art4eq-velcorearth}
v_{\rm lsr}  =v_{\odot}  + \delta v_{\odot}
\end{eqnarray}

\begin{table*}
\caption{Log of the UES observations ($R \approx 5.2 \times10^{4}$) of
HD~56126}
\label{art4tab-obs}
\centerline{\begin{tabular}{llllcl}
\hline
\multicolumn{2}{c}{Date}& Int. time& Air  & Wavelength   &$\delta v _{\oplus}$
\\
Day        & U.T.       & [sec.]   & mass & [\AA]        &  [\kms]   \\
\hline
           &            &          &      &              &       \\
24/02/1992 & 21:39      & 600      & 1.06 & 4430 -  6887 &-21.20 \\
24/02/1992 & 21:57      & 600      & 1.06 & 5409 - 10302 &-21.34 \\
           &            &          &      &              &       \\
\hline
\end{tabular}}
\end{table*}

\section{Spectral line identification}

\subsection{Introduction}

In this section we will investigate which types of spectral lines are
found in the optical spectrum of HD~56126. The spectral lines are
classified in different categories: those from molecular
species (e.g. \Ctwo~ and CN) and those from atoms. The atomic
absorption lines are subdivided  in those from the stellar
photosphere (e.g., CI, NI, and OI) and those formed by circum- or interstellar
gas (e.g., NaI~D).

\subsection{Molecular absorption lines}

The first detection of interstellar \Ctwo~ was made by
Souza and Lutz (\cite{art4souzalutz77}) and the first detection of
interstellar CN by Dunham (\cite{art4dunham}).
Since then \Ctwo~ and CN and many other molecules  have
been observed in dense
interstellar clouds in the line of sight of a dozen bright stars
(Van~Dishoeck and De Zeeuw \cite{art4dishoeckzeeuw84}).
Photospheric molecular line absorption  has been observed
by Bergeat~\etal (\cite{art4bergeatetal76}) and Barnbaum (\cite{art4barnbaum})
among others in the optical spectra of cool carbon stars.
A list of  molecular bands of a number of astrophysically
interesting molecules can be found in Wallace (\cite{art4wallace}).

We have identified the vibrational (1,0), (2,0), and (3,0)
bands of the \Ctwo~ Phillips system ($\rm A^{1}\Pi_{u}-X^{1}\Sigma^{+}_{g}$),
the (0,0)
and (1,0) \Ctwo~ Swan bands ($\rm d^{3}\Pi_{g}-a^{3}\Pi_{u}$),
and the vibrational (1,0), (2,0), (3,0), and (4,0) bands of the
``Red system'' of  CN ($\rm A^{2}\Pi-X^{2}\Sigma^{+}$)
in the optical spectrum of HD~56126.
Although we expect \Cthree~ (Hrivnak \cite{art4hrivnak})
in the optical spectrum, we were not able to detect this, since
the \Cthree~ \Cthreeband at 4050~\AA~ is not in the wavelength range
observed. The presence of \CHp~
could not be proven for the same reason.
Table~\ref{art4tab-moldet} gives an overview of the molecular bands
identified, while line identification lists of several vibrational
bands are given in App.~\ref{art4ap-mollist} (only available at CDS).

\begin{table*}
\caption{Identified molecular bands in the optical spectrum of HD~56126}
\label{art4tab-moldet}
\centerline{\begin{tabular}{lllllrl}
\hline
Molecule&System     &\multicolumn{2}{c}{Band}&\multicolumn{2}{c}{Wavelength
range}&Remark \\
        &           &         &              &\multicolumn{2}{c}{[\AA]}
  & \\
\hline
        &           &         &              &     &          &\\
\Ctwo   &Phillips   &\philband&(1,0)         &10137&10213     &\\
\Ctwo   &Phillips   &\philband&(2,0)         & 8753& 8913     &\\
\Ctwo   &Phillips   &\philband&(3,0)         & 7717& 7809     &\\
\Ctwo   &Swan       &\swanband&(0,0)         & 5100& 5166     &\\
\Ctwo   &Swan       &\swanband&(1,0)         & 4710& 4740     &\\
        &           &         &              &     &          &\\
CN      &Red system &\redband &(1,0)         & 9142& 9202     &\\
CN      &Red system &\redband &(2,0)         & 7874& 7918     &\\
CN      &Red system &\redband &(3,0)         & 6927& 6954     &\\
CN      &Red system &\redband &(4,0)         & 6190& 6225     &\\
        &           &         &              &     &          &\\
\hline
\end{tabular}}
\end{table*}

\begin{figure*}
\centerline{\hbox{\psfig{figure=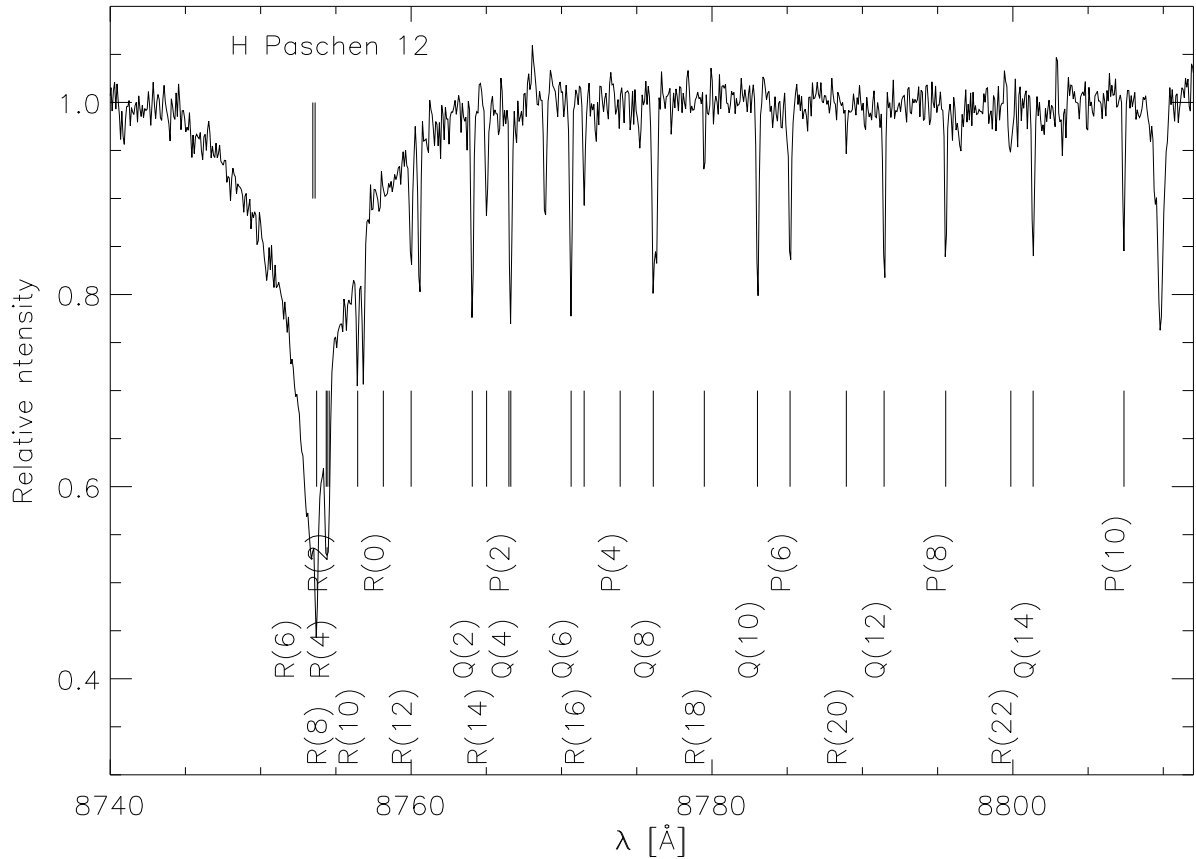,width=\textwidth}}}
\caption{The Phillips (2,0) band in the spectrum of HD~56126. The
           \Ctwo~ transitions are recognized by
           the narrow absorption profiles (not resolved).
           The photospheric absorption line at 8809.8~\AA
           is the MgI(7) transition.}
\label{art4fig-phillips20}
\end{figure*}

\begin{figure*}
\centerline{\hbox{\psfig{figure=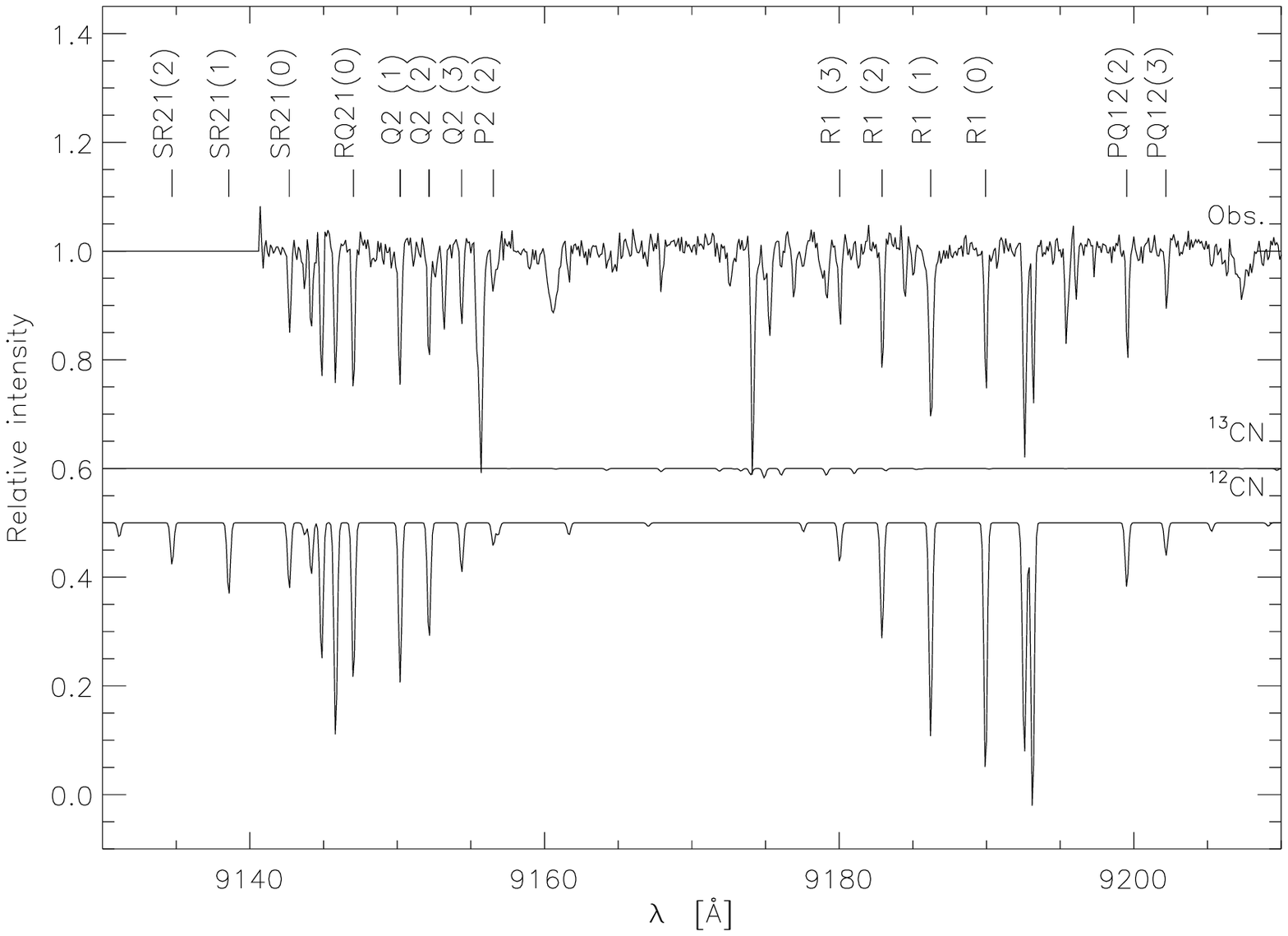,width=\textwidth}}}
\caption{The observed spectrum at the CN $\rm A^{2}\Pi-X^{2}\Sigma^{+}$ (1,0)
band.
 The spectrum has been divided by a template  spectrum to remove
the telluric lines.
A synthetic
spectrum ($T_{\rm rot}=13$~K and $N_{\rm Boltz}=13.9\times10^{14}$ cm$^{-2}$)
has
been calculated
to help to identify the
absorption lines. The $^{13}$C$^{14}$N spectrum is shifted down by
0.4 and the
$^{12}$C$^{14}$N spectrum by 0.5
($^{12}$C/$^{13}$C $=25$). The strongest (not blended)
molecular lines are marked with their identification (with $N''$ number).
The features at 9155 and 9174~\AA~
are artifacts  from the template spectrum.}
\label{art4fig-cn10}
\end{figure*}

The formation of \Ctwo~ and CN are both thought to be due to
photo-dissociation of more complex molecules by the ultraviolet radiation
field (e.g., Cherchneff~\etal \cite{art4cherglasmam}).  In the case
of \Ctwo~ the photo-dissociation reaction is given by
$\rm C_{2}H_{2} \rightarrow C_{2}H \rightarrow C_{2} $,
while in the case of CN we have the photo-dissociation reaction
$\rm HCN \rightarrow CN $.
We note that the HCN abundance in this object is quite small
(Omont~\etal \cite{art4omont}).
We have calculated the ratio between the IS UV flux (Draine \cite{art4draine})
and the stellar UV flux ($T_{\rm eff}=6500$~K Kurucz model).
At a typical distance of the AGB remnant,
$r\geq10^{16}$ cm, the stellar UV radiation field is about a
factor $10^{4}$ more intense (in the number of available UV photons)
than the interstellar UV radiation flux
for $\lambda \leq 2000$~\AA. Therefore the layers where \Ctwo~ and CN
exist are probably on both the inner and outer radius confined
by photo-dissociation processes. When the star evolves along the
post-AGB to higher temperatures, the stellar UV field increases in
intensity and the inner radius of the \Ctwo~ and CN layers will move
outwards. Due to the absence of a proper chemical model for the
circumstellar chemistry of post-AGB stars, we will use the
model for the AGB star IRC+10216 (Cherchneff~\etal \cite{art4cherglasmam}).

Molecules can exist at low
temperatures like in the photosphere of carbon stars and in interstellar
clouds ($T_{\rm gas} \leq 3000$~K).
It is obvious that these conditions do not hold in the
photosphere of HD~56126 with an effective temperature
of $T_{\rm eff}= 6500$~K. Our conclusion is thus that the molecular lines are
not photospheric but inter- or rather circumstellar.
This idea is sustained by the fact that the heliocentric
radial velocity of the molecular absorption lines  is
$-8.5\pm0.6$~\kms blue-shifted
with respect to the central velocity (system velocity of HD~56126)
derived from  the molecular
line emission (CO lines: see Table~\ref{art4tab-AGB}).

The average velocities for
different bands of different molecules are the same
(Table~\ref{art4tab-physpar}). The
only exception is the Phillips (1,0) band which has
an average velocity 6~\kms higher than the other bands. We determined
velocities of photospheric lines in the same wavelength region and
found a similar shift of 6 to 10~\kms with respect to photospheric lines
at shorter wavelengths. Since no accurate wavelength
information of the ThAr lamp was available for wavelengths longer than
9648~\AA~
at the time of reducing the spectra,
the wavelength calibration of this part of the spectrum was
obtained by extrapolation of the wavelength calibration fit. For
wavelengths longer than 6548~\AA~ the wavelength calibration is less
accurate, and the difference in velocity of the Phillips (1,0) is
due to an error in the wavelength calibration.

In this article we have identified those molecular bands which are clearly
 present in the
spectrum. But there are also bands present which can only be found by
studying a small
part of the spectrum in detail. In the process of identifying the CN (3,0)
band
another molecular band was found. So far it has not been possible to make an
identification,  but
the lines are also present in a spectrum obtained in December 1993, and they
are shifted
in wavelength. This excludes a telluric origin and favors the idea that these
lines are from circumstellar origin. Table~\ref{art4tab-unid} gives the
observed wavelength and
equivalent width of the unidentified molecular lines
with a depression larger than 5\%. The third column gives
the predicted laboratory wavelength in air assuming that the line-forming
region is the same as
for \Ctwo~ and CN.

\begin{table}
\caption{List of unidentified molecular lines}
\label{art4tab-unid}
\centerline{\begin{tabular}{llll}
\hline
$\lambda_{\rm obs}$&$W$ & $\lambda_{\rm lab}^{\rm predicted}$& remark\\
{}~[\AA]         &[m\AA]&[\AA]  &                       \\
\hline
                &       &        &                    \\
6911.7          &33.1   & 6909.4 &                    \\
6912.6          &49.8   & 6910.3 & blend (band head?) \\
6912.8          &24.9   & 6910.5 & blend              \\
6914.6          &15.1   & 6912.3 &                    \\
6915.0          &23.1   & 6912.7 &                    \\
6917.3          &25.5   & 6915.0 &                    \\
6920.2          &10.9   & 6917.8 &                    \\
6921.9          &29.7   & 6919.6 &                    \\
6924.9          &30.2   & 6922.6 &                    \\
6925.8          &18.2   & 6923.5 &                    \\
6930.8          &22.0   & 6928.5 &                    \\
                &       &        &                    \\
\hline
\end{tabular}}
\end{table}

\begin{figure}
\centerline{\hbox{\psfig{figure=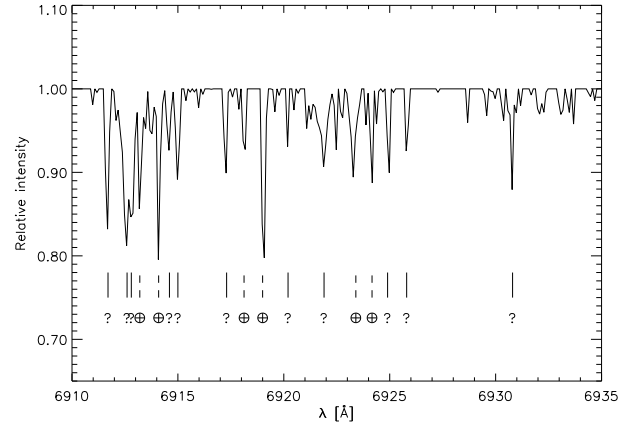,width=\columnwidth}}}
\caption{An unidentified molecular band.
The ``?'' signs denote unidentified lines, whereas
an ``$\oplus$'' denotes a telluric line which was
not completely removed by dividing
the spectrum by the reference stars HR~4049.
All lines with a depression of $\geq 5$\%
are significant.}
\label{art4fig-unident}
\end{figure}

\begin{figure*}
\centerline{\hbox{\psfig{figure=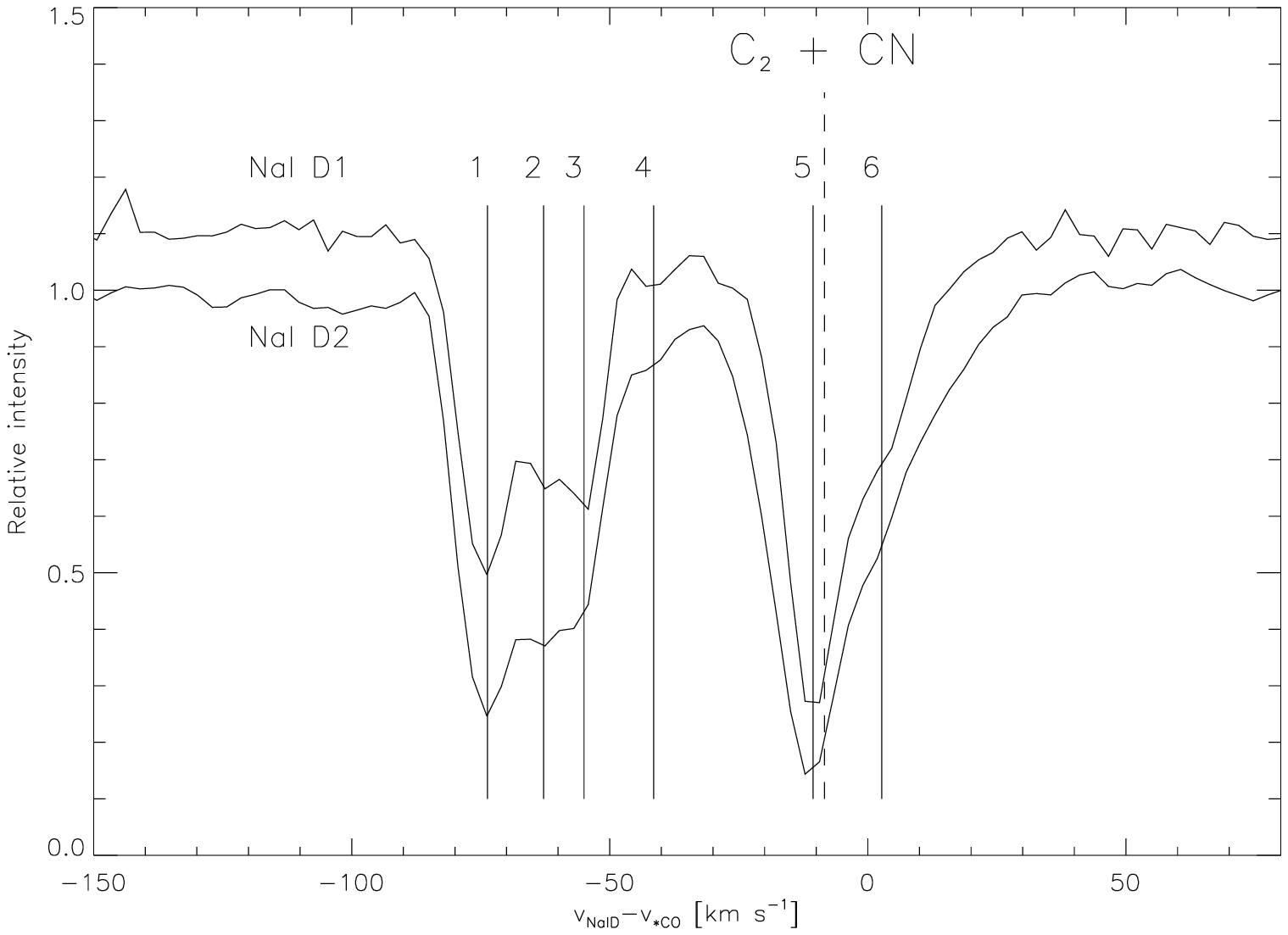,width=\textwidth}}}
\caption{The NaI~D1 (top, shifted by 0.1 in relative intensity)
           and D2 (bottom) absorption profiles show two
           well-separated broad components. The most blue-shifted
           profile is of interstellar origin and shows at least
           four different components (no. 1 to 4).
           The less blue-shifted profile
           has the large contribution from the AGB wind (no. 5). Component
           no. 6 is photospheric.}
\label{art4fig-nai}
\end{figure*}

\subsection{Atomic line absorption}

An abundance study by Klochkova (\cite{art4klochkova}) using
optical echelle spectra while assuming LTE and hydrostatic
equilibrium showed that HD~56126 has an overabundance of
CNO and the heavy elements, and a large excess of all
s-process elements. This abundance pattern is expected for
a post-AGB star which has experienced the third dredge-up on the AGB.

We have analyzed the optical photospheric spectrum and found that almost all
lines are asymmetric. A recent study by Oudmaijer and Bakker
(\cite{art4oudmayerbakker}) shows that most photospheric lines, e.g.,
H${\alpha}$,
FeI and FeII, etc., show changes in absorption line profiles on time scales
 of less than a month.
They attribute this to a pulsation of the photosphere with a period
shorter than 65 days. If the pulsation
scenario is correct, this will raise questions about the interpretation of
the abundance studies: pulsation could give an emission component in
the absorption features and thus change the observed equivalent width
in a complicated way.

\subsubsection{CI, NI, and OI absorption lines}

To allow an accurate determination of the photospheric velocity we have
fitted Gaussians to the line profile of
selected CI, NI and OI lines. As the lines are
asymmetric in shape we have fitted the profiles using the core of
the feature as the most important fitting criteria.  A list of
identification, equivalent width and depth of the lines
is presented in App.~\ref{art4ap-atomlist} (at CDS)
and the results are summarized
in Table~\ref{art4tab-velphot}.

The photospheric radial velocity thus obtained, $82.3\pm0.6$~km~s$^{-1}$,
is almost equal to the average central velocity of the CO line emission
of $85.6\pm0.5$~\kms (Nyman~\etal \cite{art4nymanetal92} and
Bujarrabal~\etal \cite{art4bujarrabaletal92}). The small difference of 3.3~\kms
is probably caused by pulsation of photosphere: the star is in a phase
of expansion.

\begin{table}
\caption{Velocities of photospheric absorption lines}
\label{art4tab-velphot}
\centerline{\begin{tabular}{llll}
\hline
Ion         &$\chi$ [eV]      & Number       & $v_{\odot}$         \\
(multiplet) &                 & of lines     & [\kms]              \\
\hline
            &                 &              &                      \\
CI          & $\geq$  7.45    &  3           & 80.7$\pm$0.8         \\
NI          & $\geq$ 10.28    &  1           & 83                   \\
OI          & $\geq$  9.11    &  3           & 83.7$\pm$0.3         \\
            &                 &              &                      \\
average     & $\geq$  7.45    &  7           & 82.3$\pm$0.6         \\
            &                 &              &                      \\
\hline
\end{tabular}}
\end{table}

\subsubsection{The absorption components in the NaI~D lines}

In view of the fact that the AGB remnant is cool (dust temperatures
of about $T_{\rm dust}=200$~K), the spectral features most likely to have a
contribution
from  circumstellar material are the
resonance absorption lines. For
this study we have chosen the NaI~D doublet.
HD~56126 has a strong infrared excess and will probably show both an
inter- and circumstellar contribution in the NaI~D lines.

A similar study has been made of the NaI~D lines profiles in the spectra of
Planetary Nebulae (Dinerstein~\etal \cite{art4dinerstein}). They have shown
that there is a circumstellar component present in the lines profiles of the
NaI~D lines.

The NaI~D1 (5895.923~\AA) and D2 (5889.953~\AA)
lines in the spectrum of HD~56126 show a very complex
structure (Fig.~\ref{art4fig-nai}).
By comparing the profile of the  D1 line with the D2 line
we can see which absorption features are real and which are artifacts.
At least six different components could be identified.
The four most blue-shifted
components form one broad feature, while the two other
components form a second less blue-shifted broad feature.
In fitting Gaussian profiles to all components with the exception of
the photospheric component (no. 6) we have determined the velocities
of the individual components in
the D1 and D2 lines independently. After extracting five Gaussian
profiles, we have assumed that the residual of the spectrum is the
photospheric contribution. The velocity of the latter contributor
is determined with an eyeball fit. A summary of the determined velocities
from the NaI~D1 and D2, and \Ctwo~ and CN lines are given in
Table~\ref{art4tab-naI}.
The error on the individual velocities is about 3~km~s$^{-1}$. The only
exception is the velocity of component 6, as this is the only component
which was isolated after subtracting of the other.
The estimated error on the velocity of component 6 is 6~km~s$^{-1}$.

\begin{table*}
\caption{Doppler (heliocentric) velocities  and equivalent width
           of the six
           components in the absorption profile of the NaI~D1 and
           D2 lines. The data  on molecular lines absorption is
           given for comparison. The average CO velocity of $85.6\pm0.5$~\kms
           has been used to determine the outflow velocity.}
\label{art4tab-naI}
\centerline{\begin{tabular}{llllllll}
\hline
no.        &
$v_{\rm D1}$   &
$W_{\rm D1}$   &
$v_{\rm D2}$   &
$W_{\rm D2}$   &
$v_{\rm NaI~D}$ &
$v_{\rm D}-v_{\rm *CO}$&
remark     \\
           &
[\kms]     &
[m\AA]     &
[\kms]     &
[m\AA]     &
[\kms]     &
[\kms]     &
          \\
\hline
            &        &          &        &          &        &               &
   \\
     1      &$12\pm3$&$154\pm5$ &$12\pm3$&$218\pm16$&$12\pm2$&$-73.8\pm2.0$  &
IS  \\
     2      &$23\pm3$&$ 55\pm5$ &$23\pm3$&$ 67\pm18$&$23\pm2$&$-62.9\pm2.0$  &
IS  \\
     3      &$30\pm3$&$ 90\pm5$ &$31\pm3$&$114\pm6 $&$31\pm2$&$-55.1\pm2.0$  &
IS  \\
     4      &$44\pm3$&$ 16\pm5$ &$44\pm3$&$ 33\pm~5$&$44\pm2$&$-41.6\pm2.0$  &
IS  \\
     5      &$75\pm3$&$240\pm5$ &$75\pm3$&$316\pm23$&$75\pm2$&$-10.7\pm2.0$  &
CS  \\
     6      &$89\pm6$&$103\pm10$&$87\pm6$&$ 87\pm10$&$88\pm4$&$ +2.6\pm2.0$  &
Ph  \\
            &        &          &        &          &        &               &
   \\
\hline
            &        &          &        &          &        &               &
   \\
            &        &          &        &          &$v$     &$v_{*}-v_{\rm
*CO}$& \\
            &        &          &        &          &[\kms]  & [\kms]        &
   \\
            &        &          &        &          &        &               &
   \\
\multicolumn{2}{l}{CO          }&&       &          &$85.6\pm0.5$&$ 0.0
$&CS  \\
\multicolumn{2}{l}{\Ctwo~ and CN}&&       &
&$77.1\pm0.3$&$-8.5\pm0.6$&CS  \\
            &       &         &          &          &      &         \\
\hline
\end{tabular}}
\noindent
\centerline{$v$: heliocentrically corrected radial velocity;
$W$: the equivalent width}
\end{table*}

\subsection{Conclusions}

At this stage we note that the Doppler
velocity found for the \Ctwo~ and CN molecules
is equal to the velocity found for the NaI Comp~5. This suggests
that they are formed in the same region. Comp~6 is very close to the
photospheric velocity. In Sect.~5 we will use
the Galactic rotation curve to
distinguish between the interstellar, circumstellar, and
photospheric contribution and argue that the NaI~D1 and D2
components 1, 2, 3, and 4 are probably interstellar. \Ctwo, CN,
and the NaI~D component no. 5 are thought to be of circumstellar
origin. The NaI~D component 6 is of photospheric origin.

\section{Physical conditions of line-forming region  of
molecular line absorption}

\subsection{Introduction}

An analysis of the spectra of molecular bands allows the deduction of
chemical composition,  rotational
temperatures, column densities, and radial velocities
of the line-forming region.  Examples of the detection of molecular
bands are shown in Figs.~\ref{art4fig-phillips20} and \ref{art4fig-cn10}.
{}From these figures we see immediately that
these molecular lines are not resolved by the instrument.
The width of the molecular absorption lines
is much smaller than the width of the observed photospheric
(e.g., CI, NI, OI) lines.

In this section we look in detail at the physical conditions
occurring in the molecular line-forming region by an analysis of the
molecular bands. Only in the case of an optically thin medium can
the rotational temperature and the column density
(assuming a Boltzmann distribution over the $J''$ levels) be derived
in a simple way. In the
optically thin case there is a linear relation between the
observed equivalent width and population density of the lower level of
the transition. Using simple
formulae this gives the rotational temperature and the column density.
In this discussion we will focus on the \Ctwo~ Phillips bands, but the
same technique can be applied to CN. Data of the CN bands was
extracted from the SCAN-CN tape of J{\o}rgenson and Larson (\cite{art4jorlar}).

In order to describe the method of deriving physical parameters from
the observations, we discuss the notation for the rotational levels and
lines of diatomic molecules  (Whiting \cite{art4whiting}).
A transition is denoted by ($v'J'$,$v''J''$) where the first index, $v'J'$,
is  the upper level and the second index, $v''J''$, the
lower level. $v$ and $J$ denote
the vibrational and  rotational  quantum number respectively.
Since we only consider the Phillips bands,
all transitions are between the same electronic states,
from ${\rm X^{1}\Sigma_{g}^{+}}$ to ${\rm A^{1}\Pi_{u} }$,
and we observe different vibrational bands ($v'$,$v''$):
(1,0), (2,0), and (3,0). Each vibrational band consists of
numerous lines from rotational transitions ($J'$,$J''$). The difference
in oscillator strength between the
vibrational bands is expressed in the band oscillator strength.
In deriving
physical parameters from the observations we will make use of
optically thin lines.
The (3,0) vibrational band of \Ctwo~ is weaker than the (1,0) and (2,0)
vibrational band. This makes the
(3,0) more suitable  for our analysis than the (2,0) and (1,0)
vibrational band.

\subsection{Rotational diagram}

\subsubsection{Case of \Ctwo}

In the optically thin  case ($\tau \ll 1$, where the absorption
lines are on the linear part of the curve of growth)
the column density for molecules
in the ($J'$,$J''$) transition is given by Eq.~\ref{art4eq-opthin}.

\begin{equation}
\label{art4eq-opthin}
N_{v''J''} {\rm ~[cm^{-2}]}  = 1.13 \times 10^{17}
\frac{W_{\lambda}^{J'J''} {\rm ~[m\AA]}}{f_{J'J''}\lambda^{2}_{J'J''}
{\rm ~[\AA]}},
\end{equation}

\noindent
with $W_{\lambda}^{J'J''}$ the observed equivalent width in m\AA,
$f_{J'J''}$ the oscillator
strength as given by Eq.~\ref{art4eq-oscillator}, and $\lambda_{J'J''}$ the
vacuum wavelength of the ($J',J''$) transition.

The oscillator strength $f_{J'J''}$ is calculated
(Eq.~\ref{art4eq-oscillator})
from the tabulated band oscillator strength $f_{v'v''}$
(Table~\ref{art4tab-molconst}), the band head
$\nu_{\rm band}$ (Table~\ref{art4tab-molconst}), the computed values
for the frequency of the transition, $\nu_{J'J''}$, and the normalized
H\"{o}nl-London factors (line intensity factor), $S_{J'J''}$.
The H\"{o}nl-London factors for the \Ctwo~ Phillips bands
were computed for the A-X system ($\Delta\Lambda=1$) . The general
formulae for the H\"{o}nl-London factors (Herzberg~\cite{art4herzberg})
are simplified to Eq.~\ref{art4eq-sp}, \ref{art4eq-sq},
and \ref{art4eq-sr} and normalized such
that for each
$J''$, $\Sigma_{J'} \frac{S_{J',J''}}{g''_{e}/g'_{e}
\left( 2 J''+1 \right)} \equiv 1 $.

\begin{eqnarray}
\label{art4eq-sp}
S^{\rm P}_{J''} = (J''-1) \\
\label{art4eq-sq}
S^{\rm Q}_{J''} = (2J''+2) \\
\label{art4eq-sr}
S^{\rm R}_{J''} = (J''+2) \\
\label{art4eq-oscillator}
f_{J'J''} = f_{v'v''} \left[
\frac{\nu_{J'J''}}{\nu_{\rm band}} \left(
                          \frac{S_{J'J''}}{g''_{e}/g'_{e}(2J''+1)}
                          \right)
                    \right],
\end{eqnarray}

\noindent
with $g''_{e}/g'_{e}$ the ratio of electronic degeneracy factor.
For the \Ctwo~ and
CN A-X system $g''_{e}/g'_{e}=2$.
Each rotational level has degeneracy of (2$J''$+1), and the energy
of the lower rotational level is given by E$_{J''}$
(Eq.~\ref{art4eq-exenergy}). It should be noted that
as \Ctwo~ is a homo-nuclear molecule without a permanent dipole
moment, symmetry of rotational states does not allow odd-numbered
rotational levels in the $X^{1}\Sigma_{g}^{+}$ electronic state to exist.
The absorption oscillator strength and laboratory wavelength in air
for the lines used in the analyses are given in App.~\ref{art4ap-mollist}
(at CDS).

If we apply Eq.~\ref{art4eq-opthin} to lines which are not optically thin,
we underestimate the column density. This is clearly demonstrated in
the computed column densities for low $J''$ values
of the Phillips (1,0) band (see App.~\ref{art4ap-mollist}).
In the case of
optically thin lines the column density derived for a given $J''$ value
should be the same for the P, Q, and R branch lines.
But in the case of optically thick lines the column density derived from
the Q  branch line is the lowest while the column density from the P
branch line is the highest (lowest  $f_{J'J''}$).

\begin{table}
\caption{Molecular parameters}
\label{art4tab-molpar}
\centerline{\begin{tabular}{lll}
\hline
Parameter               & \Ctwo     & CN         \\
                        & \philband & \redband   \\
\hline
                        &           &            \\
${\rm B_{\rm e} [cm^{-1}]}$ & 1.820101  & 1.8996     \\
                        &           &            \\
\hline
\end{tabular}}
\centerline{Herzberg \cite{art4herzberg}}
\end{table}

\begin{table}
\caption{Molecular parameters used in this work to calculate the oscillator
strength $f_{J',J''}$. Items which are not used in this work are intentionly
left blank.}
\label{art4tab-molconst}
\centerline{\begin{tabular}{llll}
\hline
band          &$\nu_{\rm band}$(vac)&$f_{v',v''}$      &ref.     \\
              &[cm$^{-1}$]       &                     &$\lambda$\\
\hline
              &                  &                     &   \\
Phillips (1,0)& 9854.0247$^{(1)}$&2.84$~10^{-3}~^{(2)}$& 1 \\
Phillips (2,0)&11413.8250$^{(1)}$&1.67$~10^{-3}~^{(2)}$& 1 \\
Phillips (3,0)&12947.81  $^{(3)}$&7.52$~10^{-4}~^{(2)}$& 1 \\
              &                  &                     &   \\
Swan (0,0)    &                  &1.5 $~10^{-3}~^{(5)}$& 4 \\
Swan (1,0)    &                  &7.6 $~10^{-3}~^{(5)}$& 4 \\
              &                  &                     &   \\
CN (1,0)      &                  &1.50$        ~^{(5)}$& 6 \\
CN (2,0)      &                  &0.76$        ~^{(5)}$& 6 \\
CN (3,0)      &                  &0.28$        ~^{(5)}$& 6 \\
CN (4,0)      &                  &                     & 6 \\
              &                  &                     &   \\
\hline
\end{tabular}}
1:  Chauville~\etal \cite{art4chauvilleetal77};
2:  Van Dishoeck \cite{art4dishoeck83};
3:  Ballik and Ramsay \cite{art4ballikramsay76};
4:  Amiot \cite{art4amiot};
5:  Davis~\etal \cite{art4davisetal};
6:  J{\o}rgenson and Larson \cite{art4jorlar}
\end{table}

\noindent
By assuming a Boltzmann distribution over the rotational levels the population
density of level $J''$ is given by

\begin{equation}
 \label{art4eq-boltzmann}
   N_{v''J''}  = \frac{N_{\rm Boltz}}{Q_{\rm r}} (2J''+1)
      \exp ^{ \frac{-E_{J''}}{kT_{\rm rot}}}
\end{equation}

\noindent
with

\begin{equation}
\label{art4eq-exenergy}
 \frac{E_{J''}}{k} = \frac{B_{\rm e}hc}{k} J'' \left( J''+1 \right)  {\rm~[K]}
\end{equation}

\noindent
and

\begin{equation}
\label{art4eq-qval}
 Q_{\rm r}= \sum_{J''}^{\infty} (2J''+1) \exp ^{ \frac{-E_{J''}}{kT_{\rm rot}}}
{}.
\end{equation}

\noindent
$B_{\rm e}$ represents the rotational constant  of the lower
electronic level (Table~\ref{art4tab-molpar}), and $k$, $h$, and
$c$  the Boltzmann, Planck constant,  and the velocity of light
respectively.

The rotational diagram with on the vertical axis
\(\ln \left( N_{J''}/(2J''+1) \right) \)  and on the horizontal axis
\( E_{J''}/k \) has a slope of \( a=-1/T_{\rm rot} \) and an offset of
 \( b=\ln \left( N_{\rm Boltz}/Q_{\rm r} \right) \). Thus the temperature is
given by the slope, and the column density is given by the
offset of the first order fit in the rotational diagram.

\begin{figure*}
\centerline{\hbox{\psfig{figure=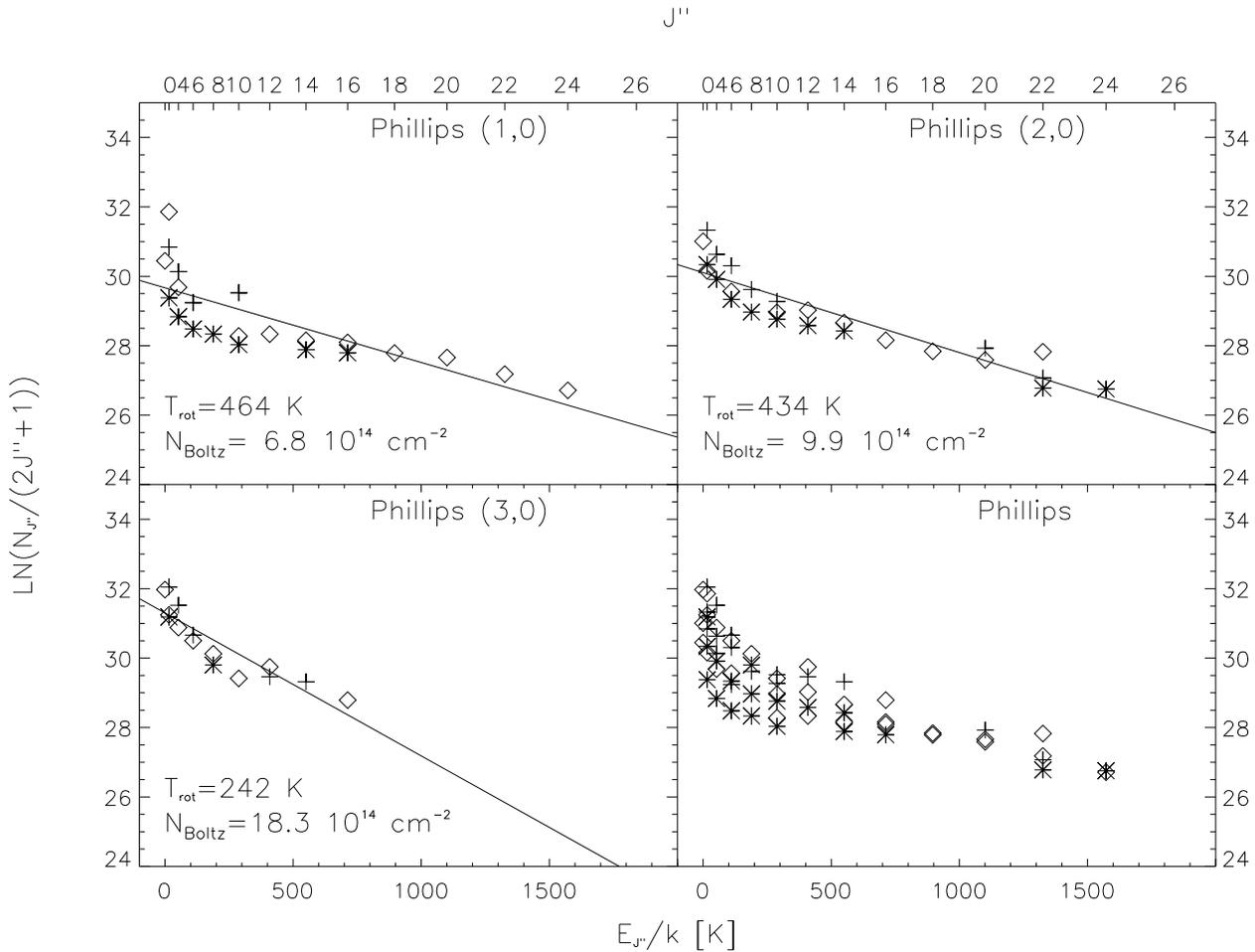,width=\textwidth}}}
\caption{Rotational diagram  of the Phillips (1,0), (2,0), and (3,0) bands.
           The P, Q, and R branch lines are denoted by a
           plus, asterisk, and square respectively. Only the weakest
           rotational band (3,0) is optically thin and
           gives an almost linear relation in the rotational diagram.}
\label{art4fig-rotphillips}
\end{figure*}

\begin{figure*}
\centerline{\hbox{\psfig{figure=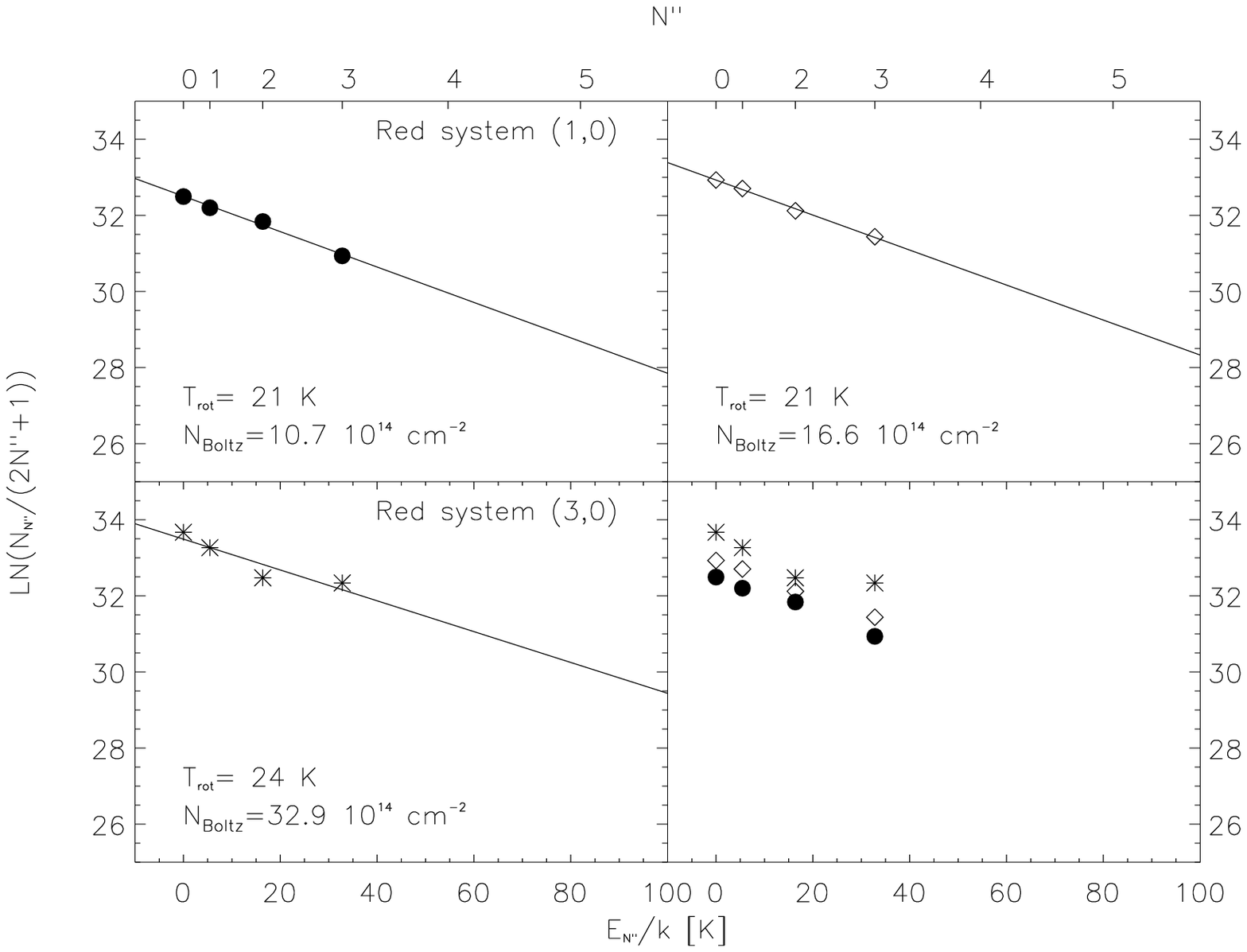,width=\textwidth}}}
\caption{Rotational diagram of the Red system of CN.}
\label{art4fig-cnrot}
\end{figure*}

The effect of optical
depth on the derived column densities can be seen in the
rotational diagrams (Fig.~\ref{art4fig-rotphillips}). In the rotational
diagram of the (1,0) band, the Q branch lines lie lower than the P branch
lines.
In the case of the (2,0) band one should note that the R branch lines
(squares) up to $J''=12$ are blended with the Paschen 12 line.
If we do not take into account R($J''=0,2,4,6,8,10,12$), then we notice that
also for the (2,0) band the Q branch lines lie lower than the P branch lines.
In the rotational diagram of the (3,0) band we do not see
a systematic difference in column density between the different
branches. Based on the derived column densities for the P, Q, and R
branch lines, we conclude that the (1,0) and (2,0) band are optically
thick while the (3,0) band is optically thin.

The population density of the $J''$ rotational
levels does not decrease monotonically with $J''$, it reaches a
maximum where the population decreases due to an increase
of rotational energy of the levels ($E_{J''}$), and
the population increases due to
an  increase of degeneracy of the level ($2J''+1$). This means
that the highest optical depth for $T_{\rm rot} \approx 242$~K
is reached around $J''=8,10,12$, and
a local minimum is observed in the rotational diagram. At low $J''$
values (e.g. $J''=0,2$) and at high $J''$ values ($J''\geq 18$) the
(1,0) and (2,0) vibrational bands are optically thin.

{}From the derived column densities of the rotation levels $J''$
of the (3,0) band,
we can compute the rotational temperature and the column density of the
vibration level $v''$. In doing this we assume that the
population distribution over the rotational levels follows a
Boltzmann distribution.
Since the rotational diagram of the optically thin vibrational band
(3,0) gives a straight line, the approximation is to first order allowed.
It should be noted that this does not necessarily mean that
$T_{\rm kin}=T_{\rm rot}$ as the excitation of the molecule is due to the
combined effect of collisional and radiative (de-)excitation.
In the case of a homo-nuclear specie like \Ctwo~ $T_{\rm rot} \geq T_{\rm
kin}$,
while for a hetero-nuclear specie like CN  $T_{\rm rot} \leq T_{\rm kin}$.
Detailed modeling is required to find $T_{\rm kin}$ from the molecular
excitation.

\subsubsection{Case of CN}

In the case of CN the spin angular momentum $S$ is decoupled
from the total angular momentum and the energy levels are
described by the total angular momentum quantum number $N$.
Eq.~\ref{art4eq-opthin}  is valid for CN when $J''$ is replaced
by $N''J''$. Then $N_{v''N''}$ is given by $N_{v''N''J''}$ summed over
the $J''=N''-1/2$ and $J''=N''+1/2$ levels
 (Table~\ref{art4tab-cn}).
Eqs.~\ref{art4eq-boltzmann}, \ref{art4eq-exenergy},
and \ref{art4eq-qval} are also valid when $J''$ is replaced by
$N''$.

\begin{table}
\caption{The observed column densities (in units of $10^{12}$
cm$^{-2}$) of the CN Red system}
\label{art4tab-cn}
\centerline{\begin{tabular}{lllll}
\hline
$N''$ & $J''$            &(1,0)&(2,0)&(3,0)\\
\hline
      &                  &     &     &     \\
0     & 1/2              & 130 & 200 & 420 \\
1     & 1/2              & 116 & 240 & 328 \\
1     & 3/2              & 174 & ?   & 412 \\
2     & 3/2              & 176 & 225 & 256 \\
2     & 5/2              & 160 & 221 & 377 \\
3     & 5/2              &  81 & 158 & 388 \\
3     & 7/2              & 110 & ?   & ?   \\
      &                  &     &     &     \\
\hline
\end{tabular}}
\end{table}

The calculation of the oscillator strength for CN using
an equation analogue to
Eqs.~\ref{art4eq-sp}, \ref{art4eq-sq}, \ref{art4eq-sr},
and \ref{art4eq-oscillator} is not trivial, and we have therefore
adopted the $\log gf$ values for CN of the SCAN-CN tape.
The $\log gf$ values have been multiplied by a factor of 0.734
as suggested by the authors of the tape. For a detailed description
of CN and further reading we refer to
J{\o}rgenson and Larson (\cite{art4jorlar}).

\subsection{Determination of $T_{\rm rot}$, $N_{\rm Boltz}$ and $N_{\rm
observed}$}

Applying the Boltzmann law to derive a rotational temperature
and column density of \Ctwo~
(Fig.~\ref{art4fig-rotphillips})
yields values ranging from $T_{\rm rot}=242$~K to 464~K,
and \Ctwo~ column densities
of $N_{\rm Boltz}=0.7~{\rm to}~1.8\times10^{15} {\rm ~cm^{-2}}$.
These results are sensitive
to the molecular lines and molecular bands
included in the determination of the rotational temperature
and the column density. The reason for this is that this method only
gives valid results if the lines are optically thin and
if the population density of the rotational levels can be described
by a Boltzmann distribution.

As the (3,0) band is the only optically thin band and
is most reliable for determining the rotational temperature and column
densities, the values derived from this band will be adopted
for the line-forming region: $T_{\rm rot} = 242\pm 20$~K,
$N_{\rm Boltz}=1.8 \times 10^{15}$ cm$^{-2}$.  Using the same technique on the
CN (3,0) band we find $T_{\rm rot}=24\pm5$~K and
$N_{\rm Boltz} = 3.3 \times 10^{15}$ cm$^{-2}$.
This gives a particle abundance ratio of $f_{\rm C_{2}}/f_{\rm CN}=0.5$
in reasonable agreement
with the predicted molecular peak
abundances ratio for the carbon-rich envelope of
IRC~+10216  of $f_{\rm C_{2}}/f_{\rm CN}=1.0$
(Cherchneff and Glassgold \cite{art4cherglas}).
A second measure of the column density can be obtained by adding the
observed
column densities of each energy level. In that case we find for
\Ctwo~ (3,0) $N_{\rm obs}= 2.0 \times 10^{15}$ cm$^{-2}$ and for CN (3,0)
$N_{\rm obs} = 2.7 \times 10^{15}$ cm$^{-2}$. This is reasonably in agreement
in
view of the fact that by adding the observed column densities
only a limited number of levels are used, and that some of the
observed lines will be optically thick.

A rotational temperature of 242~K for \Ctwo~ is significantly higher than
expected for interstellar molecules ($T_{\rm rot} \leq 100$~K). The
line-forming
region is hotter than expected if it were in radiative equilibrium with
the interstellar radiation field, and this suggests that the molecules are
in radiative equilibrium with the stellar radiation field.
This is consistent with the velocity difference of $-8.5\pm0.6$~\kms
found  between the millimeter line emission (CO) and the optical
molecular line absorption and suggests that the line-forming region
is circumstellar.
This
velocity difference is interpreted as the molecular outflow velocity.
The fact that we only observe one velocity for the optically
thin and optically thick lines of \Ctwo~ and CN suggests that the
line-forming region is of constant velocity and that there is no significant
velocity gradient.

\begin{table*}
\caption{Physical parameters derived from molecular bands}
\label{art4tab-physpar}
\centerline{\begin{tabular}{lllllll}
\hline
Band& & &$v_{\odot}$&$N_{\rm obs}$          &$N_{\rm Boltz}$
&$T_{\rm rot}$\\
    & & &[\kms]     &\multicolumn{2}{c}{[10$^{14}$~cm$^{-2}$]}   &[K] \\
\hline
               &         &            &              &     &     &   \\
\Ctwo~Phillips &\philband&(1,0)       &$(82.7\pm0.4)$& 9.9 & 6.8 &484\\
\Ctwo~Phillips &\philband&(2,0)       &$ 76.8\pm0.2 $&12.3 & 9.9 &434\\
\Ctwo~Phillips &\philband&(3,0)       &$ 76.7\pm0.2 $&20.0 &18.3 &242\\
\Ctwo~Swan     &\swanband&(0,0)       &              &     &     &   \\
\Ctwo~Swan     &\swanband&(1,0)       &              &     &     &   \\
               &         &            &              &     &     &   \\
CN~Red system  &\redband&(1,0)        &$ 77.7\pm0.2 $& 9.5 &10.7 & 21\\
CN~Red system  &\redband&(2,0)        &$ 76.3\pm0.4 $&11.4 &16.6 & 21\\
CN~Red system  &\redband&(3,0)        &$ 78.0\pm0.2 $&26.7 &32.9 & 24\\
CN~Red system  &\redband&(4,0)        &              &     &     &   \\
               &        &             &              &     &     &   \\
\multicolumn{3}{r}{Average velocity}  &$ 77.1\pm0.3 $&     &     &   \\
               &        &             &              &     &     &   \\
\hline
\end{tabular}}
\end{table*}

\section{Predicted velocities of interstellar absorption lines}
\label{art4sec-islines}

In the former sections we have seen that the Doppler velocity
of the molecular line absorption and the NaI~D lines are not
at the system velocity, but are from inter- or circumstellar origin.
In this section we look at the predicted Doppler velocities for
interstellar contributions. Using the simple argument that
HD~56126 is located
above the Galactic plane ($b=10^{\rm o}$) with a relatively high
heliocentric radial velocity of $85.6\pm0.5$~km~s$^{-1}$, one expects to
be able to distinguish interstellar from circumstellar contributions
on the basis of observed velocities.
We can estimate which observed Doppler velocities are probably
not of interstellar origin.

{}From model evolutionary tracks (Sch\"onberner \cite{art4schonberner83})
one predicts for a typical post-AGB star with M$_{\rm
post-AGB}=0.6$~M$_{\odot}$, a
M$_{\rm bol}=-4.30$. From the observations (Table~\ref{art4tab-star})
we find m$_{\rm bol}=6.34$, which places
the star at a distance of 1.3~kpc and 0.2~kpc above the Galactic plane.

For the velocity structure of the Galactic disk we adopt the simple
approximation given in Lang (\cite{art4lang}) and subtract a correction
term to convert to heliocentric velocities (Eq.~\ref{art4eq-galrot}).
Inserting the  Galactic coordinates of HD~56126 gives a formula which
expresses the heliocentric radial velocity of the Galactic disk as
 a function of distance (Eq.~\ref{art4eq-velis}).

\begin{equation}
\label{art4eq-galrot}
   v_{\odot}({\rm IS})=
   -\delta v_{\odot} + A~D~\sin 2l~\cos^{2} b ~[{\rm~km~s^{-1}}]
\end{equation}

\noindent
with the following parameters:

\[\delta v_{\odot}= -14.11 {\rm~km~s^{-1}} \]
\[ A=15 {\rm ~km~s^{-1}~kpc^{-1}}  \]
\[ (l;b)_{\rm HD~56126} = (206.75^{\rm o};9.99^{\rm o}) \]

\noindent
gives:

\begin{equation}
 \label{art4eq-velis}
 v_{\odot}({\rm IS}) =
            14.11 + 11.69 \times D [{\rm kpc}]~~[{\rm~km~s^{-1}}] \hfill
\end{equation}

Eq.~\ref{art4eq-velis} gives a lower- and upper-limit
on the predicted velocities of interstellar lines using
a distance of  $D=0$~kpc and $D=1.3$~kpc respectively. We find a velocity
range between 14 and  30~km~s$^{-1}$. That the lower limit is not zero is
of course due to the motion of the sun with respect to
the solar neighborhood.
The upper limit of 30~\kms is also the velocity expected
for HD~56126 if it is of population I origin.  The fact that
there is a difference of 56~\kms between ``predicted'' and observed velocity
is the reason why HD~56126 is classified as a high-velocity star.

We can now look at Table~\ref{art4tab-naI} and distinguish the absorption
components in the NaI~D lines in two categories. We see that only components
2 and 3 are within the range expected for IS lines. As the predicted range
of IS velocities is only a rough estimate, we argue that the whole
absorption feature between
heliocentric velocities of 10 and 44~\kms (components 1,2,3, and 4)
is probably of interstellar origin. This means that there are at least four
independent
interstellar clouds in the line of sight to HD~56126.

We can now immediately conclude that the absorption features
with heliocentric velocities between
74 and 89~\kms (component 5 and 6) are not of interstellar origin,
and we attribute them to circumstellar and photospheric material.

The velocity found for component 5 of the NaI~D complex
is equal to the velocity found
for molecular lines absorption
(\Ctwo~ and CN) which strongly suggests that these lines
are formed in the same region. In Sect.~4 we have discussed the physical
conditions of the molecular line forming region and found
rotational temperatures in agreement with a circumstellar origin.
We are now able to conclude that the NaI~D component 5 and the
\Ctwo~ and CN molecular absorption lines are from circumstellar origin.
The photospheric contribution is estimated using the standard star
$\alpha$~Per
(F5I). This star has NaI~D1 and D2 equivalent width of 540 and 500~m\AA.
The NaI~D component 6
has an equivalent width of the same order, and we argue that this component
is photospheric.

\section{Discussion}

The primary aim of this study is to make a line identification of
the molecular bands found in the optical spectrum of HD~56126 and
to identify the line-forming region.
By looking at the absorption
components in the NaI~D1 and D2 lines and a simple model for the Galactic
rotation curve, we are able to separate the NaI~D1 and D2
interstellar component (1, 2, 3, and 4)
from the circumstellar component (5, C$_{2}$, and CN).
The rotational temperature of super-thermal \Ctwo~ is $T_{\rm rot}=242\pm20$~K,
while for sub-thermal CN we find $T_{\rm rot}=24\pm5$~K.

In the post-AGB phase the main constituent of the circumstellar material
is the last mass-loss episode the star suffered on the AGB.
A small contribution of the circumstellar matter might be
post-AGB mass loss, but
because the escape velocity of a post-AGB object is much higher than
for an AGB object, the wind velocities are also much higher.
The presence of molecular absorption lines at an expansion velocity
of $-8.5$~\kms is too low to be attributed to post-AGB mass-loss.
This means that these molecules are located in the AGB remnant.

In this study we add to the tracers of the AGB remnant, i.e., infrared energy
distribution and millimeter line emission (e.g., CO) a new one: molecular
absorption lines.
Combining information from different tracers allows us to derive
time scales and mass-loss rates. Table~\ref{art4tab-AGB} gives an overview
of the information on the AGB remnant derived from infrared, millimeter,
and optical data. Furthermore, we predict the presence of  \Htwo~
(e.g., 1108~\AA),
\Ctwo~ (e.g., 1341~\AA), and CO (e.g., 1509~\AA) in the UV.

Due to self-shielding, CO is not easily photo-dissociated
by the UV radiation field and therefore is
present out to a large distance  from the star in the
circumstellar environment.  This is not the
case for \Ctwo~ and CN, as they are much more easily photo-dissociated by
the UV radiation field ($\lambda \leq 2000$~\AA).
The circumstellar layer where \Ctwo~ and CN
are present is determined by an interplay between the stellar
and interstellar UV radiation field and the local particle density.
This implies that
CO is detected over the whole AGB remnant, while \Ctwo~ and CN are formed
over a limited distance range. Furthermore,
the millimeter
CO emission lines are biased to line-forming regions with a large volume
(large distance from the star), while absorption lines are biased
to high densities (close to the star). The infrared excess
is biased to dust with high temperatures, and thus also to
material close to the star.
As a first  approximation to derive time scales and mass-loss rates
from the detection of molecular absorption lines we will assume that
the molecular absorption lines originate between
the dust inner and outer radius.

\begin{figure}
\centerline{ \hbox{\psfig{figure=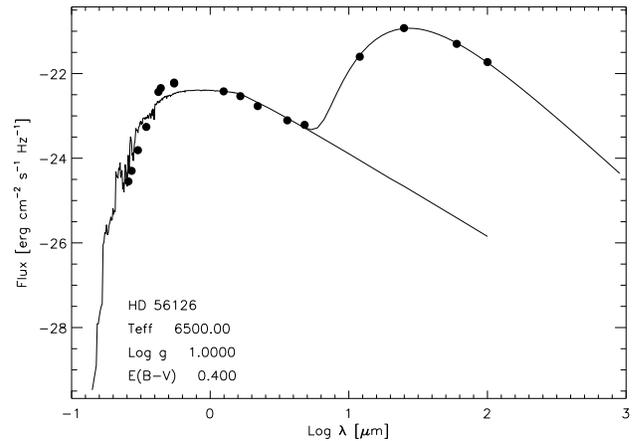,width=\columnwidth}}}
\caption{The spectral energy distribution of HD~56126 fitted to
a $T_{\rm eff}=6500$~K, $\log g=1.0$ Kurucz model
and an optically thin dust model. The
difference in UV flux between model and observations is probably
due to an anomalous circumstellar reddening law.}
\label{art4fig-sed}
\end{figure}

We have fitted an optically thin dust model (Waters~\etal
\cite{art4watcotgeb}) to the spectral energy distribution
(Fig.~\ref{art4fig-sed}). The best fit was reached for
a detached dust shell with a dust inner radius of
$r_{1}=2\times10^{3}$~R$_{*}$ ($=7 \times 10^{15}$ cm),
a dust outer radius of $r_{2} = 8\times10^{3}$~R$_{*}$, and a dust temperature
of $T_{\rm dust}= 190-110$~K. This dust inner radius is in agreement with
the work by Hrivnak~\etal (\cite{art4hrivnaketal89}). To calculated
the AGB mass-loss rate from the IR,
an estimate of the reddening is needed. Here
we discuss two approaches to estimated the reddening: (i) from
the observed $(B-V)$ and the intrinsic $(B-V)_{0}$ we compute
$E(B-V)=0.60$ (Table~\ref{art4tab-star}). Normalizing the
model on the observed energy distribution in the near-infrared,
we find that the model strongly underestimated the observed flux near the
Balmer jump and argue that $E(B-V)=0.60$ cannot explain the observations.
(ii) for a spherically symmetric dust shell the energy radiated
in the infrared is  provided by
circumstellar reddening of starlight. We have dereddened the stellar energy
distribution and found that the infrared energy output is supported by a
circumstellar reddening of $E(B-V)=0.40$. The corresponding model
(Fig.~\ref{art4fig-sed})
is in reasonable agreement with the observations and yields
a mass-loss rate of
$\dot M_{\rm IR} = 4\times10^{-5}$~\msol~\pyr. This is about a factor eight
higher than derived by Hrivnak~\etal (\cite{art4hrivnaketal89}).
This difference in derived mass-loss rate can partly be explained
by the  differences in grain characteristics, stellar luminosity,  and
density law adopted in the two different models. The deficiency of
UV flux might indicate the presence of an additional
anomalous circumstellar reddening (e.g., $\lambda^{-1}$ for
$\lambda \leq 0.4~\mu$m).

\begin{table*}
\caption{An overview of the observational tracers of the AGB remnant.
         A distance of 1.3~kpc derived for an 0.565~\msol post-AGB model
         (R$_{*}=50$~R$_{\odot}$, $T_{\rm eff}=6500$~K, $\log L =3.57$~\lsol),
         and $\log g=0.5$ has been assumed in this table.}
\label{art4tab-AGB}
\centerline{\begin{tabular}{llll}
\hline
Origin            &Parameters                               &$v_{\rm
exp}$&Ref.\\
                  &                                         &[\kms]       &
\\
\hline
                  &                                         &             & \\
Dust              &$T_{\rm dust}=110-190$~K                 &             &1\\
                  &$r_{1}=2\times10^{3}$~R$_{\ast}$         &             &1\\
                  &$r_{2}=8\times10^{3}$~R$_{\ast}$         &             &1\\
                  &$\log \dot M_{\rm IR}=-4.4\pm1.0$~\msol~\pyr&
&1\\
                  &$r_{1}=1.4\times10^{3}$~R$_{\ast}$       &             &2\\
                  &$\log \dot M_{\rm IR}=-5.3\pm1.0$~\msol~\pyr&
&2\\
$^{12}$CO($J=1-0$)&$v=84.5\pm0.5$~\kms                      &$-12.0\pm1.0$&3\\
$^{12}$CO($J=2-1$)&$v=87  \pm0.5$~\kms                      &             &4\\
$^{12}$CO($J=1-0$)&$v=86  \pm0.5$~\kms                      &             &4\\
$^{13}$CO($J=2-1$)&$v=85  \pm0.5$~\kms                      &             &4\\
$^{12}$CO($J=1-0$)&$v=85.5\pm1.0$~\kms                      &$-10.0\pm2.0$&5\\
Average           &$v_{\rm *CO}=85.6\pm0.5$~\kms            &             &1\\
                  &$\log \dot M_{\rm CO}=-4.92\pm0.10$~\msol~\pyr&
&6\\
\Ctwo             &$T_{\rm rot}=242\pm20$~K                 &             &1\\
                  &$\log N(\Ctwo)=15.3\pm0.3$~cm$^{-2}$     &             &1\\
CN                &$T_{\rm rot}=24\pm5$~K                   &             &1\\
                  &$\log N(CN)=15.5\pm0.3$~cm$^{-2}$        &             &1\\
\Ctwo~ and CN     &$v=77.1\pm0.3$~\kms                      &$-8.5\pm0.6$ &1\\
                  &$\log \dot M_{\rm mol}=-5\pm1$~\msol~\pyr&             &1\\
NaI~D1 and D2     &$v_{\rm NaI~D}=75\pm2$~\kms              &$-11\pm2$    &1\\
                  &                                         &             & \\
\hline
\end{tabular}}
1: this study;
2: Hrivnak~\etal \cite{art4hrivnaketal89};
3: Nyman~\etal \cite{art4nymanetal92};
4: Bujarrabal~\etal \cite{art4bujarrabaletal92};
5: Zuckerman~\etal \cite{art4zuckermanetal86};
6: Omont~\etal \cite{art4omont}
\end{table*}

Combining information from the energy distribution
(inner dust radius) and the molecular
absorption lines (outflow velocities)
makes it possible to calculate the elapsed time since HD~56126 has left the
AGB (Eq.~\ref{art4eq-time}).

\begin{equation}
\label{art4eq-time}
t_{\rm left~AGB}
=\frac{2.2\times10^{-2} . r_{1} [R_{\ast}]}{v_{\rm exp}
{\rm ~[km~s^{-1}]}}
= 250 {\rm ~years}
\end{equation}

We find that HD~56126 left the AGB about 250 years ago. Such a short
time scale is not unusual for post-AGB evolution, as theoretical
evolutionary tracks predict transition times from AGB to Planetary
Nebula as short as 10,000 years (Sch\"{o}nberner \cite{art4schonberner83}).
It is also consistent with the change of spectral type from G5 to F5 in the
last 50 years.

\begin{eqnarray}
\label{art4eq-mdot}
\dot M = 9\times10^{-6} .
                \left( \frac{r_{2}r_{1}/\left(r_{2}-r_{1}) \right)
                 }{9.3\times10^{15}~{\rm cm}} \right) .
                \left( \frac{v_{\rm exp}}{8.5~{\rm km~s^{-1}} } \right) .
 \nonumber \\
                \left( \frac{N_{\rm mol}   }{2.0\times10^{15}~{\rm cm^{-2}}}
                \right) .
                \left( \frac{2\times10^{-6}}{f_{\rm mol}   } \right)
                ~\left[ {\rm M_{\odot}~yr^{-1}} \right]
\end{eqnarray}

\noindent
The mass-loss rate from the molecules can be estimated
with Eq.~\ref{art4eq-mdot}, with the distance $r$ in cm, $v_{\rm exp}$ in
km~s$^{-1}$,
$N_{\rm mol}$ the column density in cm$^{-2}$
(\Ctwo : $N_{\rm mol}=N_{\rm obs}$ (3,0); CN: $N_{\rm mol}=N_{\rm Boltz}$
(3,0)),
and $f_{\rm mol}$ the
molecular abundance relative to \Htwo.
We assume that the molecular absorption lines
are formed between the dust inner radius ($r_{1}=2\times10^{3}R_{*}$)
and dust outer radius ($r_{2}=8\times10^{3}R_{*}$).  Peak abundances
of $f_{\rm C_{2}}=f_{\rm CN}=2\times10^{-6}$ are
taken from the work by Cherchneff and Glassgold (\cite{art4cherglas}) in
which they model
the carbon-rich circumstellar envelope of IRC+10216. The mass-loss
rates found are $\dot M_{\rm C_{2}}=9\times10^{-6}$~\msol~\pyr~ and
$\dot M_{\rm CN}=14\times10^{-6}$~\msol~\pyr,
and we will adopt a mass-loss rate of
$\dot M_{\rm mol}=1\times10^{-5}$~\msol~\pyr~ from the molecular
absorption lines with an estimated error of one order.
With the above calculation (Eq.~\ref{art4eq-mdot}) we want above all
to show that the detection of molecular lines absorption allows the
determination of the AGB mass-loss rate. But at this stage the uncertainties
in the calculation due to the molecular abundances and the exact distance
of the line-forming  region to the central star lead to a rather large
uncertainty in the value for the derived mass-loss rate. Proper chemical
models are required to solve this problem.

An independent technique to determine the mass-loss rate is modeling the
population distribution over $J''$ levels of the electronic-vibrational
ground level of \Ctwo~
(Bakker~\etal \cite{art4bakkeredin}; Van Dishoeck and Black
\cite{art4dishoeckblack82}). This allows the determination of the
number of collisional partners (H and H$_{2}$) and hence the
AGB mass-loss rate. This work is still in progress and will
be published in a separate paper.

The problem of a change of mass-loss rate along the AGB can only
be studied if there are tracers available which are formed in different
regions of the AGB remnant.
The observed millimeter CO emission is at larger distance from the
star than the \Ctwo~ and CN molecular lines. Comparing the expansion
velocities of the \Ctwo~ and CN absorption lines with the
CO expansion velocities tells us whether there is a change in
expansion velocity of the AGB wind during the last episode of the
AGB phase. As CO has an expansion velocity of $-10\pm2$~\kms whereas
\Ctwo~ and CN have an expansion velocity of $-8.5\pm0.6$~km~s$^{-1}$, the
difference between those velocities of 2.6~\kms is only
a 2$\sigma$ deviation, and we
are not able to show whether or not the expansion velocity of the
AGB wind has changed as the star evolved along the AGB.

\section{Conclusion}

We have identified molecular line absorption in the optical spectrum
of the post-AGB star HD~56126 of \Ctwo~ (Phillips (1,0), (2,0),
and (3,0); Swan (1,0) and (0,0)), and CN (Red system (1,0), (2,0),
(3,0), and (4,0)).
For all molecular lines we find the same expansion velocity
relative to the CO millimeter line emission of
$-8.5\pm0.6$~\kms and identify the line-forming region
in the AGB remnant.
We find that the \Ctwo~ molecular lines in the (1,0) and (2,0) Phillips band
are optically thick, whereas the Phillips (3,0) band is optically thin.
For the \Ctwo~  Phillips (3,0) we find a
rotational temperature of $T_{\rm rot}=242\pm20$~K and a column density of
$N_{\rm Boltz} = 2.0\times10^{15}$~cm$^{-2}$.
For the CN Red system (3,0) we find $T_{\rm rot}=24\pm5$~K and
$N_{\rm Boltz} = 3.3\times10^{15}$~cm$^{-2}$.
Assuming that the molecular absorption lines are formed at the dust inner
radius and the molecular abundances are equal to those in the circumstellar
envelope of IRC+10216, we find a mass-loss rate of
$\dot M_{\rm mol}=1\times10^{-5}$~\msol~\pyr~
with an estimated error of one order.

We have started a systematic survey of
molecular absorption and emission lines in the optical spectra
of post-AGB stars in order to fully exploit this new approach
to study  the AGB remnant.

\acknowledgements{
This work would not have been possible without the contribution of
Ton Schoenmaker and Ewine van Dishoeck.
In reducing the raw data to handsome spectra Ton
provides us with the data to do this study whereas Ewine has
introduced us in the field of molecular spectroscopy and
astrochemistry.
The authors want to thank Christoffel Waelkens, Ren\'{e} Oudmaijer,
and Hans van Winckel
for the stimulating and constructive discussions on this work.
The referee, Prof. A. Omont, has made a valuable contribution by
suggesting important changes to this article.
EJB was supported by grant no. 782-371-040 by ASTRON,
which receives funds from the Netherlands Organization for
the Advancement of Pure Research (NWO).
LBFMW is supported by the Royal Netherlands Academy of Arts and Sciences.
This research has made use of the Simbad database, operated at
CDS, Strasbourg, France.}

\appendix

\section{Molecular lines identification of \Ctwo~ and CN}
\label{art4ap-mollist}

We present a complete line identification of
the \Ctwo~ Phillips bands (1,0), (2,0), and (3,0), and
the CN Red system (1,0) and (2,0). Each table (only available
at CDS) gives for
a rotation band the line identification (column 1) and the
laboratory wavelength (column 2) in {\bf air}. The wavelength
conversion from vacuum to air was made by applying Cauchy's formula.
The data on the laboratory wavelength and the oscillator strength
were obtained from a variety of sources. Most of the references
are given in Table~\ref{art4tab-molpar}.
The observational parameters of the rotational line identified
are given in columns 3 and 4, the observed wavelength and
the observed equivalent width (m\AA)  respectively. From these values the
column densities can be calculated under the assumption of
optically thin lines. The conversion from equivalent width to
column density is explained in Sect.~4. The last column gives a
remark if applicable. It should be noted that we do not have
a complete wavelength coverage in the wavelength interval
observed. For longer wavelength than 6030~\AA~
the echellogram is larger in the dispersion direction
than the size of the CCD. This results in gaps between
two successive orders in the spectrum.
The data of CN has been taken from J{\o}rgenson and Jarson (\cite{art4jorlar}).

\section{Photospheric CI, NI, and OI lines}
\label{art4ap-atomlist}

The heliocentric velocity of the star have been obtained by
an accurate Gaussian fit to high-excitation CI, NI, and OI lines.
The table gives a list of identified lines and the
observational parameters. This table has the same layout as
the tables presented in App.~\ref{art4ap-mollist}.
For a description of the table
we refer to the introduction of  App.~\ref{art4ap-mollist}.
The atomic data is taken from Moore (\cite{art4moore72}).

\begin{table}
\caption{Line identification list of photospheric CI, NI, and OI}
\label{art4tab-CNO}
\begin{center}
\begin{tabular}{|lllllll|}
\hline
$\lambda_{\rm lab}$&Atom  &$\chi$&$\lambda_{\rm obs}$&$W_{\lambda}$&$D$
&$v_{\odot}$\\
% leave this line blank, otherwise it will give an error
{}~[\AA]           &(mtp) & [eV]      &[\AA]  &[m\AA]  &\%   &[km/s]\\
\hline
9078.32        &CI(3)& 7.45 &9081.4 &626.8   &0.74 & 80     \\
9061.48        &CI(3)& 7.45 &9064.6 &525.5   &0.58 & 82     \\
9062.53        &CI(3)& 7.45 &9065.5 &683.4   &0.62 & 80     \\
8242.34        &NI(2)&10.29 &8245.2 & 59.2   &0.13 & 83     \\
7771.96        &OI(1)& 9.11 &7774.7 &570.6   &0.59 & 84     \\
7774.18        &OI(1)& 9.11 &7776.9 &544.8   &0.59 & 84     \\
7775.40        &OI(1)& 9.11 &7778.1 &474.9   &0.56 & 83     \\
\hline
\end{tabular}
\end{center}
\end{table}

% \end{document}

\begin{table*}
\caption{Identification of the (1,0) band in the
  $ {\rm C_{2}~ A^{1}\Pi_{u}-X^{1}\Sigma_{g}^{+} }$ Phillips system}
\label{art4tab-phillips10}
\begin{small}
\centerline{\begin{tabular}{|lllllll|}
\hline
Branch ($J''$)  & $\lambda_{\rm lab}$ & $\lambda_{\rm obs}$
& $W_{\lambda}$ & $f_{J'J''}10^{4}$&$N_{J''}$   & Remark \\
      & [\AA]     & [\AA]    &[m\AA] &        & [$10^{12}$~cm$^{-2}$]&
\\
\hline
R(~0) & 10143.726 & 10147.18 & ~43.2 & 28.41  & ~16.7 & blend on red wing \\
P(~2) & 10154.900 & 10158.42 & ~32.2 & ~2.84  & 124.2 &                    \\
Q(~2) & 10148.354 & 10151.89 & ~37.2 & 14.20  & ~28.7 &                    \\
R(~2) & 10138.542 & 10141.97 & ~35.5 & 11.36  & 342.3 &                    \\
P(~4) & 10164.764 & 10168.22 & ~47.5 & ~4.72  & 110.1 & cosmic hit        \\
Q(~4) & 10151.525 & 10154.99 & ~38.9 & 14.19  & ~30.1 & cosmic hit        \\
R(~4) & 10135.151 & 10138.61 & ~60.5 & ~9.48  & ~70.2 &                   \\
P(~6) & 10176.254 & 10179.81 & ~32.4 & ~5.45  & ~64.9 &                   \\
Q(~6) & 10156.518 & 10160.05 & ~39.4 & 14.18  & ~30.4 &                   \\
R(~6) & 10133.605 & 10137.21 & ~87.3 & ~8.75  &       & blended with R(~8)\\
P(~8) & 10189.695 &          &       & ~5.82  &       & $\scriptstyle \leq $14
\AA \\
Q(~8) & 10163.326 & 10166.89 & ~44.5 & 14.18  & ~34.3 &                    \\
R(~8) & 10133.857 & 10137.21 & ~87.3 & ~8.36  &       & blended with R(~6) \\
P(10) & 10205.000 & 10208.57 & ~45.8 & ~3.56  & 139.6 &                    \\
Q(10) & 10171.965 & 10175.57 & ~40.7 & 14.16  & ~31.4 &                    \\
R(10) & 10135.925 & 10139.39 & ~29.7 & ~8.12  & ~40.2 &                   \\
P(12) & 10222.174 &          &       & ~6.20  &       & not observed       \\
Q(12) & 10182.436 &          &       & 14.15  &       & $\scriptstyle \leq $15
\AA \\
R(12) & 10139.807 & 10143.33 & ~36.6 & ~7.96  & ~50.5 &                    \\
P(14) & 10241.248 &          &       & ~6.31  &       & not observed       \\
Q(14) & 10194.757 & 10198.38 & ~48.6 & 14.13  & ~18.9 &                    \\
R(14) & 10145.507 & 10148.98 & ~35.0 & ~7.83  & ~49.1 &                    \\
P(16) & 10262.232 &          &       & ~6.38  &       & not observed       \\
Q(16) & 10208.933 & 10212.52 & ~50.5 & 14.11  & ~38.8 &                    \\
R(16) & 10153.034 & 10156.60 & ~37.1 & ~7.74  & ~52.5 &                    \\
P(18) & 10285.144 &          &       & ~6.44  &       & not observed       \\
Q(18) & 10224.985 &          &       & 14.09  &       & not observed       \\
R(18) & 10162.400 & 10166.00 & ~30.3 & ~7.66  & ~43.3 &                    \\
P(20) & 10310.011 &          &       & ~6.48  &       & $\scriptsize \leq$15
\AA \\
Q(20) & 10242.923 &          &       & 14.07  &       & not observed       \\
R(20) & 10173.608 & 10177.19 & ~29.4 & ~7.60  & ~42.2 &                    \\
P(22) & 10336.863 &          &       & ~6.50  &       & $\scriptsize \leq$ 15
\AA \\
Q(22) & 10262.768 &          &       & 14.04  &       & not Observed       \\
R(22) & 10186.670 & 10190.27 & ~19.9 & ~7.54  & ~28.7 &                    \\
P(24) & 10365.711 &          &       & ~6.52  &       & $\scriptsize \leq $15
\AA\\
Q(24) & 10284.541 &          &       & 14.01  &       & not observed       \\
R(24) & 10201.603 & 10205.08 & ~13.5 & ~7.49  & ~19.6 &                    \\
P(26) & 10396.593 &          &       & ~6.54  &       & $\scriptsize \leq$ 15
\AA \\
Q(26) & 10308.256 &          &       & 13.98  &       & $\scriptsize \leq$ 15
\AA \\
R(26) & 10218.418 &          &       & ~7.45  &       & $\scriptsize \leq$ 15
\AA \\
P(28) & 10429.547 &          &       & ~6.54  &       & not observed       \\
Q(28) & 10333.947 &          &       & 13.94  &       & not observed       \\
R(28) & 10237.138 &          &       & ~7.41  &       & not observed       \\
P(30) & 10464.598 &          &       & ~6.54  &       & not observed       \\
Q(30) & 10361.633 &          &       & 13.90  &       & $\scriptsize \leq 15$
\AA \\
R(30) & 10257.781 &          &       & ~7.37  &       & not observed       \\
%P(32) & 10501.780 &          &       & ~6.54  &       & not observed       \\
%Q(32) & 10391.343 &          &       & 13.86  &       & $\scriptsize \leq 15$
%%\AA\\
%R(32) & 10280.362 &          &       & ~7.33  &       & not observed       \\
%P(34) & 10541.139 &          &       & ~6.54  &       & not observed       \\
%Q(34) & 10423.101 &          &       & 13.82  &       & not observed       \\
%R(34) & 10304.910 &          &       & ~7.29  &       & $\scriptsize \leq 15$
%%\AA\\
%P(36) & 10582.708 &          &       & ~6.53  &       & not observed       \\
%Q(36) & 10457.358 &          &       & 13.78  &       & not observed       \\
%R(36) & 10331.448 &          &       & ~7.26  &       & $\scriptsize \leq 15$
%%\AA\\
\hline
\end{tabular}}
\end{small}
\end{table*}

\begin{table*}
\caption{Identification of the (2,0) band in the
  $ {\rm C_{2}~ A^{1}\Pi_{u}-X^{1}\Sigma_{g}^{+} }$ Phillips system}
\label{art4tab-phillips20}
\begin{small}
\centerline{\begin{tabular}{|lllllll|}
\hline
Branch ($J''$)  & $\lambda_{\rm lab}$ & $\lambda_{\rm obs}$
& $W_{\lambda}$ & $f_{J'J''}10^{4}$ & $N_{J''}$ & Remark \\
      & [\AA]    & [\AA]   &[m\AA] &        & [$10^{12}$ cm$^{-2}$]&     \\
\hline
R(~0) & 8757.686 & 8760.54 & ~33.2 & 16.70  & ~29.3 &                    \\
P(~2) & 8766.031 & 8768.90 & ~22.9 & ~1.67  & 201.6 &                    \\
Q(~2) & 8761.197 & 8764.08 & ~42.5 & ~8.35  & ~74.9 &                    \\
R(~2) & 8753.949 & 8756.81 & ~28.4 & ~6.68  & ~44.6 &                    \\
P(~4) & 8773.430 & 8776.30 & ~34.3 & ~2.78  & 181.1 &                    \\
Q(~4) & 8763.754 & 8766.62 & ~50.0 & ~8.35  & ~88.1 &                    \\
R(~4) & 8751.687 & 8754.40 & ~37.2 & ~5.57  &       & blended with R(~8) \\
P(~6) & 8782.311 & 8785.18 & ~41.1 & ~3.20  & 188.2 &                    \\
Q(~6) & 8767.762 & 8770.61 & ~40.6 & ~8.34  & ~71.6 &                    \\
R(~6) & 8750.850 & 8753.70 & ~31.3 & ~5.14  & ~52.0 & blended with P(12) \\
P(~8) & 8792.652 & 8795.53 & ~29.1 & ~3.43  & 124.0 &                    \\
Q(~8) & 8773.223 & 8776.09 & ~36.9 & ~8.34  & ~65.0 &                    \\
R(~8) & 8751.490 & 8754.40 & ~37.2 & ~4.92  &       & blended with R(~4) \\
P(10) & 8804.502 & 8807.37 & ~26.5 & ~3.56  & 108.5 &                    \\
Q(10) & 8780.144 & 8783.04 & ~37.0 & ~8.33  & ~65.2 &                    \\
R(10) & 8753.581 & 8756.43 & ~25.7 & ~4.77  & ~45.1 &                    \\
P(12) & 8817.830 &         &       & ~3.65  &       & not observed       \\
Q(12) & 8788.561 & 8791.44 & ~36.7 & ~8.32  & ~64.5 &                    \\
R(12) & 8757.130 & 8759.99 & ~32.0 & ~4.68  & 100.8 &                    \\
P(14) & 8832.682 &         &       & ~3.71  &       & not observed       \\
Q(14) & 8798.462 & 8801.34 & ~30.4 & ~8.31  & ~53.4 &                    \\
R(14) & 8762.147 & 8765.04 & ~25.6 & ~4.61  & ~81.7 &                    \\
P(16) & 8849.075 &         &       & ~3.76  &       & not observed       \\
Q(16) & 8809.844 &         &       & ~8.30  &       & not observed       \\
R(16) & 8768.631 & 8771.52 & ~17.3 & ~4.55  & ~55.9 &                    \\
P(18) & 8866.997 &         &       & ~3.79  &       & not observed       \\
Q(18) & 8822.728 &         &       & ~8.29  &       & not observed       \\
R(18) & 8776.611 & 8779.49 & ~14.0 & ~4.50  &~ 29.7 &                    \\
P(20) & 8886.487 & 8889.43 & ~14.7 & ~3.81  & ~55.2 &                    \\
Q(20) & 8837.122 &         &       & ~8.28  &       & not observed       \\
R(20) & 8786.050 & 8788.94 & ~12.0 & ~4.47  & ~24.7 &                    \\
P(22) & 8907.545 & 8910.47 & ~~6.9 & ~3.83  & ~25.7 &                    \\
Q(22) & 8853.044 & 8855.96 & ~11.0 & ~8.26  & ~19.2 &                    \\
R(22) & 8796.979 & 8799.85 &       & ~4.43  & ~54.7 & tentatively detected\\
P(24) & 8930.172 &         &       & ~3.84  &       & $\scriptsize \leq 5$ \AA
\\
Q(24) & 8870.519 & 8873.38 & ~11.7 & ~8.24  & ~20.4 &                    \\
R(24) & 8809.410 &         &       & ~4.41  &       & not observed       \\
P(26) & 8954.440 &         &       & ~3.85  &       & not observed       \\
Q(26) & 8889.535 & 8892.43 & ~10.8 & ~8.23  &       & tentative detection\\
R(26) & 8823.379 &         &       & ~4.38  &       & not observed       \\
P(28) & 8980.336 &         &       & ~3.86  &       & not observed       \\
Q(28) & 8910.135 & 8913.09 & ~~5.1 & ~8.21  &       & tentative detection\\
R(28) & 8838.846 &         &       & ~4.36  &       & not observed       \\
P(30) & 9007.897 &         &       & ~3.86  &       & $\scriptsize \leq 5$
\AA\\
Q(30) & 8932.333 &         &       & ~8.19  &       & $\scriptsize \leq 5$ \AA
\\
R(30) & 8855.876 &         &       & ~4.33  &       & $\scriptsize \leq 5$ \AA
\\
%P(32) & 9037.150 &         &       & ~3.86  &       & $\scriptsize \leq 5$ \AA
%%\\
%Q(32) & 8956.139 &         &       & ~8.17  &       & not observed       \\
%R(32) & 8874.466 &         &       & ~4.31  &       & $\scriptsize \leq 5$ \AA
%%\\
%P(34) & 9068.086 &         &       & ~3.86  &       & $\scriptsize \leq 5$ \AA
%% \\
%Q(34) & 8981.586 &         &       & ~8.14  &       & $\scriptsize \leq 5$ \AA
%%\\
%R(34) & 8894.623 &         &       & ~4.29  &       & $\scriptsize \leq 5$ \AA
%%\\
%P(36) & 9100.823 &         &       & ~3.85  &       & $\scriptsize \leq 5$ \AA
%%\\
%Q(36) & 9008.705 &         &       & ~8.12  &       & $\scriptsize \leq 5$ \AA
%%\\
%R(36) & 8916.400 &         &       & ~4.27  &       & $\scriptsize \leq 5$ \AA
%%\\
\hline
\end{tabular}}
\end{small}
\end{table*}

\begin{table*}
\caption{Identification of the (3,0) band in the
  $ {\rm C_{2}~ A^{1}\Pi_{u}-X^{1}\Sigma_{g}^{+} }$ Phillips system}
\label{art4tab-phillips30}
\begin{small}
\centerline{\begin{tabular}{|lllllll|}
\hline
Branch ($J''$)  & $\lambda_{\rm lab}$ & $\lambda_{\rm obs}$
& $W_{\lambda}$ & $f_{J'J''}10^{4}$ & $N_{J''}$& Remark \\
      & [\AA]    & [\AA]   &[m\AA]&      &[$10^{12}$ cm$^{-2}$]   &   \\
\hline
R(~0) & 7719.331 & 7721.87 & 30.5 & 7.52 & ~76.9 &                    \\
P(~2) & 7725.822 & 7728.37 & 16.4 & 0.75 & 414.0 &                    \\
Q(~2) & 7722.096 & 7724.62 & 34.7 & 3.76 & 174.9 &                    \\
R(~2) & 7716.531 & 7719.05 & 29.5 & 3.01 & 186.0 &                    \\
P(~4) & 7731.666 & 7734.21 & 29.2 & 1.25 & 441.6 &                    \\
Q(~4) & 7724.222 & 7726.76 & 35.6 & 3.76 &       & telluric pollution \\
R(~4) & 7714.947 & 7717.48 & 30.7 & 2.51 & 232.2 &                    \\
P(~6) & 7738.740 & 7741.28 & 20.6 & 1.44 & 269.9 &                    \\
Q(~6) & 7727.560 & 7730.12 & 37.9 & 3.76 &       & telluric pollution \\
R(~6) & 7714.578 & 7717.12 & 28.0 & 2.32 & 229.2 &                    \\
P(~8) & 7747.040 &         &      & 1.54 &       & not observed       \\
Q(~8) & 7732.120 & 7734.66 & 29.6 & 3.75 & 149.2 &                    \\
R(~8) & 7715.418 & 7717.94 & 24.0 & 2.21 & 206.1 &                    \\
P(10) & 7756.585 &         &      & 1.60 &       & not observed       \\
Q(10) & 7737.908 & 7740.43 & 27.9 & 3.75 &       & blended with R(18) \\
R(10) & 7717.473 & 7720.00 & 14.2 & 2.15 & 125.3 &                    \\
P(12) & 7767.372 & 7769.89 & 13.7 & 1.64 & 156.5 &                    \\
Q(12) & 7744.903 &         &      & 3.75 &       & not observed       \\
R(12) & 7720.751 & 7723.28 & 23.1 & 2.11 & 207.5 &                    \\
P(14) & 7779.431 & 7781.95 & 14.0 & 1.67 & 156.5 &                    \\
Q(14) & 7753.144 &         &      & 3.74 &       & not observed       \\
R(14) & 7725.243 & 7727.74 & 13.1 & 2.07 &       & telluric pollution \\
P(16) & 7792.741 & 7795.24 & ~6.4 & 1.69 &       & tentative detection \\
Q(16) & 7762.626 &         &      & 3.74 &       & not observed       \\
R(16) & 7730.966 & 7733.51 & 11.4 & 2.05 & 105.1 &                    \\
P(18) & 7807.322 & 7809.86 & ~3.0 & 1.71 &       & tentative detection\\
Q(18) & 7773.358 & 7775.92 & ~7.9 & 3.73 &       & blended with OI(1) \\
R(18) & 7737.908 & 7740.43 & 27.9 & 2.03 &       & blended with Q(10) \\
P(20) & 7823.201 &         &      & 1.72 &       & $\scriptsize \leq 5$ \AA \\
Q(20) & 7785.350 & 7787.85 & 11.5 & 3.73 &       & telluric pollution  \\
R(20) & 7746.098 &         &      & 2.01 &       & not Observed        \\
P(22) & 7840.387 &         &      & 1.73 &       & pollution \\
Q(22) & 7798.613 &         &      & 3.72 &       & telluric pollution  \\
R(22) & 7755.532 &         &      & 2.00 &       & not observed        \\
P(24) & 7858.889 &         &      & 1.73 &       & not observed        \\
Q(24) & 7813.138 &         &      & 3.72 &       & $\scriptsize \leq 5$ \AA  \\
R(24) & 7766.226 &         &      & 1.98 &       & $\scriptsize \leq 5$ \AA  \\
P(26) & 7878.721 &         &      & 1.74 &       & pollution  \\
Q(26) & 7828.979 &         &      & 3.71 &       & $\scriptsize \leq 5$ \AA \\
R(26) & 7778.178 &         &      & 1.97 &       & $\scriptsize \leq 5$ \AA \\
P(28) & 7899.907 &         &      & 1.74 &       & $\scriptsize \leq 5$ \AA \\
Q(28) & 7846.122 &         &      & 3.70 &       & not observed        \\
R(28) & 7791.405 &         &      & 1.96 &       & $\scriptsize \leq 5$ \AA \\
P(30) & 7922.464 &         &      & 1.74 &       & $\scriptsize \leq 5$ \AA  \\
Q(30) & 7864.595 &         &      & 3.69 &       & not observed        \\
R(30) & 7805.908 &         &      & 1.95 &       & $\scriptsize \leq 5$ \AA \\
%P(32) & 7946.406 &         &      & 1.74 &       & $\scriptsize \leq 5$ \AA \\
%Q(32) & 7884.394 &         &      & 3.68 &       & $\scriptsize \leq 5$ \AA \\
%R(32) & 7821.720 &         &      & 1.94 &       & $\scriptsize \leq 5$ \AA \\
%P(34) & 7971.783 &         &      & 1.74 &       & $\scriptsize \leq 5$ \AA
%%\\
%Q(34) & 7905.561 &         &      & 3.67 &       & $\scriptsize \leq 5$ \AA
%%\\
%R(34) & 7838.850 &         &      & 1.93 &       & $\scriptsize \leq 5$ \AA
%%\\
%P(36) & 7998.571 &         &      & 1.74 &       & $\scriptsize \leq 5$ \AA \\
%Q(36) & 7928.100 &         &      & 3.66 &       & $\scriptsize \leq 5$ \AA
%%\\
%R(36) & 7857.314 &         &      & 1.92 &       & $\scriptsize \leq 5$ \AA
%%\\
\hline
\end{tabular}}
\end{small}
\end{table*}

\begin{table*}
\caption{Identification of the (1,0) band of the
  $ {\rm CN~ A^{2}\Pi-X^{2}\Sigma^{-} }$ transition}
\label{art4tab-CN10}
\begin{small}
\centerline{\begin{tabular}{|lllllllll|}
\hline
Transition     & $J''$ & $N''$ & $\lambda_{\rm lab}$ & $\lambda_{\rm obs}$
& $W_{\lambda}$ & $f_{J'J''}10^{4} $ & $N_{N''J''}$ & Remark \\
      &      &   & [\AA]    & [\AA]   &[m\AA]&      & [$10^{12}$ cm$^{-2}$]&
    \\
\hline
SR21  &  3/2 & 1 & 9135.582 & NO      &      & 1.74 &      &  \\
SR12  &  1/2 & 0 & 9139.686 & 9142.70 & 33.7 & 2.54 & 179.4&  \\
RQ21  &  7/2 & 3 & 9141.189 & 9144.21 & 35.3 & 3.65 & 130.7&  \\
 R2   &  5/2 & 3 & 9141.169 & blend   &      & 3.43 &      &  \\
RQ21  &  5/2 & 2 & 9141.886 & 9144.88 & 57.8 & 3.99 & 195.7&  \\
 R2   &  3/2 & 2 & 9141.870 & blend   &      & 3.44 &      &  \\
RQ21  &  3/2 & 1 & 9142.838 & 9145.82 & 57.8 & 6.14 & 127.2& blended \\
 R2   &  1/2 & 1 & 9142.828 & blend   &      & 3.65 &      &  \\
RQ21  &  1/2 & 0 & 9144.042 & 9147.04 & 57.8 & 6.14 & 127.1&  \\
 Q2   &  1/2 & 1 & 9147.208 & 9150.18 & 61.5 & 6.14 & 135.2&  \\
QP12  &  3/2 & 1 & 9147.217 & blend   &      & 1.55 &      &  \\
 Q2   &  3/2 & 2 & 9149.167 & 9152.19 & 47.0 & 5.36 & 118.3&  \\
QP21  &  5/2 & 2 & 9149.182 & blend   &      & 1.74 &      &  \\
 Q2   &  5/2 & 3 & 9151.381 & 9154.38 & 31.1 & 5.50 &  76.2&  \\
QP21  &  7/2 & 3 & 9151.401 & blend   &      & 1.73 &      &  \\
 P2   &  3/2 & 2 & 9153.532 & 9156.62 & 27.4 & 1.55 & 238.2&  \\
 R1   &  7/2 & 3 & 9177.032 & 9180.04 & 31.5 & 4.71 &  89.7&  \\
 R1   &  5/2 & 2 & 9179.911 & 9182.95 & 47.6 & 5.16 & 123.6&  \\
 R1   &  3/2 & 1 & 9183.209 & 9186.24 &101.8 & 6.19 & 220.2&  \\
 R1   &  1/2 & 0 & 9186.932 & 9189.96 & 59.9 & 9.79 &  81.9&  \\
 Q1   &  7/2 & 3 & 9189.486 & 9192.61 & 97.8 & 5.44 &      & blended  \\
 Q2   &  5/2 & 2 & 9189.597 & blend   &      & 4.99 &      &  \\
QR12  &  5/2 & 3 & 9189.465 & blend   &      & 3.60 &      &  \\
QR12  &  3/2 & 2 & 9189.581 & blend   &      & 4.87 &      &  \\
QR12  &  1/2 & 1 & 9190.120 & 9193.18 & 62.9 & 8.68 &  96.9&  \\
 Q1   &  3/2 & 1 & 9190.129 & blend   &      & 4.13 &      &  \\
PQ12  &  3/2 & 2 & 9196.511 & 9199.58 & 42.2 & 3.26 & 172.8&  \\
 P1   &  5/2 & 2 & 9196.525 & blend   &      & 0.72 &      &  \\
PQ12  &  5/2 & 3 & 9199.171 & 9202.23 & 22.2 & 3.48 &  85.1&  \\
 P1   &  7/2 & 3 & 9199.192 & blend   &      & 1.24 &      &  \\
\hline
\end{tabular}}
\end{small}
\end{table*}

\begin{table*}
\caption{Identification of the (2,0) band of the
  $ {\rm CN~ A^{2}\Pi-X^{2}\Sigma^{-} }$ transition}
\label{art4tab-CN20}
\begin{small}
\centerline{\begin{tabular}{|lllllllll|}
\hline
Transition     & $J''$ & $N''$ & $\lambda_{\rm lab}$ & $\lambda_{\rm obs}$
& $W_{\lambda}$ & $f_{J'J''}10^{4}$ & $N_{N''J''}$         & Remark \\
     &       &   & [\AA]    & [\AA]   &[m\AA]&      & [$10^{12}$ cm$^{-2}$]&
    \\
\hline
SR21 &   1/2 & 0 & 7871.654 & 7874.2  & 29.4 & 1.28 &     & blended       \\
 R2  &   5/2 & 3 & 7872.889 &         &      & 1.73 &     & blended       \\
RQ21 &   7/2 & 3 & 7872.905 & 7875.4  & 18.5 & 1.84 &     & blended       \\
 R2  &   3/2 & 2 & 7873.332 &         &      & 1.73 &     & blended       \\
RQ21 &   5/2 & 2 & 7873.343 & 7875.9  & 38.8 & 2.02 &     & blended       \\
 R2  &   1/2 & 1 & 7873.985 &         &      & 1.84 &     & blended       \\
RQ21 &   3/2 & 1 & 7873.992 & 7876.5  & 45.8 & 2.26 &     & blended       \\
RQ21 &   1/2 & 0 & 7874.852 & 7877.4  & 39.1 & 3.09 &230.2&               \\
 Q2  &   1/2 & 1 & 7877.198 & 7879.8  & 98.2 & 3.09 &     & polluted      \\
QP21 &   3/2 & 1 & 7877.205 &         &      & 0.78 &     & not detected  \\
 Q2  &   3/2 & 2 & 7878.686 & 7881.2  & 35.6 & 2.70 &240.0&               \\
QP21 &   5/2 & 2 & 7878.697 &         &      & 0.88 &     & not detected  \\
 Q2  &   5/2 & 3 & 7880.384 & 7882.9  & 19.2 & 2.77 &126.2&               \\
QP21 &   7/2 & 3 & 7880.400 &         &      & 0.88 &     & not detected  \\
 P2  &   3/2 & 2 & 7881.889 &         &      & 0.78 &     & not detected  \\
 P2  &   5/2 & 3 & 7885.725 &         &      & 0.98 &     & not detected  \\
 R1  &   7/2 & 3 & 7899.481 & 7901.8  & 42.0 & 2.37 &     & pollution \\
 R1  &   5/2 & 2 & 7901.520 & 7904.1  & 31.6 & 2.59 &220.8&               \\
 R1  &   3/2 & 1 & 7903.892 & 7906.5  & 41.3 & 3.12 &239.5&               \\
 R1  &   1/2 & 0 & 7906.598 & 7909.2  & 46.3 & 4.93 &169.6&               \\
QR12 &   5/2 & 3 & 7908.597 &         &      & 1.81 &     & blended       \\
QR12 &   3/2 & 2 & 7908.611 & 7911.2  & 50.1 & 2.45 &     & blended       \\
 Q1  &   7/2 & 3 & 7908.613 &         &      & 2.73 &     & blended       \\
 Q1  &   5/2 & 2 & 7908.622 &         &      & 2.50 &     & blended       \\
QR12 &   1/2 & 1 & 7908.959 & 7911.5  & 53.3 & 4.37 &     & blended       \\
 Q1  &   3/2 & 1 & 7908.966 &         &      & 2.08 &     & blended       \\
PQ12 &   3/2 & 2 & 7913.692 & 7916.3  & 19.1 & 1.64 &210.0&               \\
 P1  &   5/2 & 2 & 7913.704 &         &      & 0.36 &     & not detected  \\
PQ12 &   5/2 & 3 & 7915.714 & 7918.3  & 18.5 & 1.75 &190.3&               \\
 P1  &   7/2 & 3 & 7915.729 &         &      & 0.62 &     & not detected  \\
OP12 &   5/2 & 3 & 7920.804 &         &      & 0.26 &     & not detected  \\
\hline
\end{tabular}}
\end{small}
\end{table*}

\begin{table*}
\caption{Identification of the (3,0) band of the
  $ {\rm CN~ A^{2}\Pi-X^{2}\Sigma^{-} }$ transition}
\label{art4tab-CN30}
\begin{small}
\centerline{\begin{tabular}{|lllllllll|}
\hline
Transition     & $J''$ & $N''$ & $\lambda_{\rm lab}$ & $\lambda_{\rm obs}$
& $W_{\lambda}$ & $f_{J'J''}10^{4}$ & $N_{N''J''}$   &Remark \\
     &       &   &[\AA]     &[\AA]    &[m\AA]&       &[$10^{12}$ cm$^{-2}$]&
   \\
\hline
SR21 &   7/2 & 3 & 6918.598 &         &      &0.3773 &     &
\\
SR21 &   5/2 & 2 & 6920.501 &         &      &0.4182 &     &
\\
SR21 &   3/2 & 1 & 6922.587 & 6924.9  & 30.2 &0.4744 &     &           weak
\\
SR21 &   1/2 & 0 & 6924.855 &         &      &0.5612 &     &
\\
 R2  &   5/2 & 3 & 6925.917 &         &      &0.7582 &     &
\\
RQ21 &   7/2 & 3 & 6925.929 &         &      &0.8108 &     &
\\
 R2  &   3/2 & 2 & 6926.197 &         &      &0.7601 &     &
\\
RQ21 &   5/2 & 2 & 6926.206 & 6928.5  & 17.2 &0.8840 &458.0&
\\
 R2  &   1/2 & 1 & 6926.659 &         &      &0.8061 &     &            blended
\\
RQ21 &   3/2 & 1 & 6926.665 & 6928.9  & 21.6 &0.9896 &     &            blended
\\
RQ21 &   1/2 & 0 & 6927.302 & 6929.6  & 27.1 &1.3570 &469.9&
\\
 Q2  &   1/2 & 1 & 6929.119 & 6931.4  & 18.9 &1.3570 &327.6&
\\
QP21 &   3/2 & 1 & 6929.124 &         &      &0.3418 &     &
\\
 Q2  &   3/2 & 2 & 6930.297 & 6932.6  & 12.9 &1.1835 &256.3&
\\
QP21 &   5/2 & 2 & 6930.305 &         &      &0.3842 &     &
\\
 Q2  &   5/2 & 3 & 6931.655 &         &      &1.2138 &     &
\\
QP21 &   7/2 & 3 & 6931.667 &         &      &0.3850 &     &
\\
 P2  &   3/2 & 2 & 6932.748 &         &      &0.3418 &     &
\\
 P2  &   5/2 & 3 & 6935.742 &         &      &0.4318 &     &
\\
 R1  &   7/2 & 3 & 6946.486 &         &      &1.0377 &     &
\\
 R1  &   5/2 & 2 & 6947.992 & 6950.3  & 14.3 &1.1356 &294.6&
\\
 R1  &   3/2 & 1 & 6949.770 & 6952.1  & 24.0 &1.3620 &412.0&
\\
 R1  &   1/2 & 0 & 6951.821 & 6954.1  & 34.1 &2.1529 &370.1&
\\
QR12 &   3/2 & 2 & 6953.418 &         &      &1.0763 &     &            blended
\\
 Q1  &   5/2 & 2 & 6953.427 &         &      &1.0982 &     &            blended
\\
QR12 &   5/2 & 3 & 6953.462 &         &      &0.7990 &     &            blended
\\
 Q1  &   7/2 & 3 & 6953.475 &         &      &1.1983 &     &            blended
\\
QR12 &   1/2 & 1 & 6953.646 & 6955.9  & 52.8 &1.9182 &     &            blended
\\
 Q1  &   3/2 & 1 & 6953.652 &         &      &0.9131 &     &            blended
\\
PQ12 &   3/2 & 2 & 6957.305 &  6959.6 &  7.9 &0.7193 &256.2&
\\
 P1  &   5/2 & 2 & 6957.314 &         &      &0.1598 &     &
\\
PQ12 &   5/2 & 3 & 6958.908 &  6961.2 & 12.8 &0.7684 &388.4&
\\
 P1  &   7/2 & 3 & 6958.920 &         &      &0.2703 &     &
\\
OP12 &   5/2 & 3 & 6962.801 &         &      &0.1122 &     &
\\
\hline
\end{tabular}}
\end{small}
\end{table*}

\end{document}